\documentclass{aa}  

\usepackage{graphicx}
\usepackage{txfonts}
\usepackage{url}
\usepackage{natbib}
\bibpunct{(}{)}{;}{a}{}{,}
\usepackage{enumitem}
\usepackage{array}

\newcommand{\fermi}{\textit{Fermi}} 

\begin{document}

   \title{A parsec-scale wobbling jet in the high-synchrotron peaked blazar PG\,1553+113.}

  \author{R.~Lico\inst{1,2}\fnmsep\thanks{Email: rlico@mpifr.de}, J.~Liu\inst{1}, M.~Giroletti\inst{2}, M.~Orienti\inst{2,3}, J.~L.~G\'omez\inst{4}, B.~G.~Piner\inst{5}, N.~R.~MacDonald\inst{1}, F.~D'Ammando\inst{2}, A.~Fuentes\inst{4}.} 

   \institute{Max-Planck-Institut f\"{u}r Radioastronomie, Auf dem H\"{u}gel 69, D-53121 Bonn, Germany
\and INAF - Istituto di Radioastronomia, via Gobetti 101, 40129 Bologna, Italy.
\and Dipartimento di Fisica e Astronomia, Universit\`a di Bologna, via Gobetti 93/3, 40129 Bologna, Italy.
\and Instituto de Astrof\'{\i}sica de Andaluc\'{\i}a, IAA-CSIC, Apdo. 3004, 18080 Granada, Spain.
\and Department of Physics and Astronomy, Whittier College, 13406 E. Philadelphia Street, Whittier, CA 90608, USA.
              }

   \date{Received ...; accepted 25 December 2019}

  \abstract
   {The detection of quasi-periodic variability in active galactic nuclei (AGNs) in general, and in blazars in particular, is key to our understanding of the origin and nature of these objects as well as their cosmological evolution. PG\,1553+113 is the first blazar showing an approximately two-year quasi-periodic pattern in its $\gamma$-ray light curve, which is also revealed at optical frequencies.
}  
   {Such quasi-periodicity might have a geometrical origin, possibly related to the precessing nature of the jet, or could be intrinsic to the source and related to pulsational accretion flow instabilities. In this work we investigate and characterise the high-resolution radio emission properties of PG\,1553+113 on parsec scales in order to differentiate between these different physical scenarios. }
   {We monitored the source with the very long baseline array (VLBA) at 15, 24, and 43\,GHz during an entire cycle of $\gamma$-ray activity in the period 2015--2017, with a cadence of about 2 months, both in total and polarised intensity. We constrained the jet position angle across the different observing epochs by investigating the total intensity ridge lines.}
   {We find a core-dominated source with a limb-brightened jet structure extending for $\sim1.5$ mas in the northeast direction whose position angle varies in time in the range $\sim40^{\circ} - 60^{\circ}$. No clear periodic pattern can be recognized in the VLBA light curves during 2015--2017 or in the 15\,GHz Owens Valley Radio Observatory light
curve during the period 2008--2018. The core region polarisation percentage varies in the range $\sim1-4\%$, and the polarisation angle varies from being roughly parallel to roughly transverse to the jet axis. We estimate a rotation measure value in the core region of $\sim-1.0 \pm 0.4 \times 10^4$ rad \ m$^{-2}$. The brightness temperature ($T_B$) is found to decrease as the frequency increases with an intrinsic value of $\sim1.5 \times 10^{10}$\,K and the estimated Doppler factor is $\sim 1.4$.
    }
   {Although the jet wobbling motion indicates that geometrical effects can produce an enhanced emission through the Doppler boosting modulation, additional mechanisms are required in order to account for the quasi-periodic variability patterns observed in $\gamma$ rays. The intrinsic $T_B$ value indicates that the total energy in the core region is dominated by the magnetic field. 
   }

\authorrunning{R. Lico et al. 2019}
\titlerunning{Pc-scale radio properties of the TeV blazar PG\,1553+113.}

   \maketitle


\section{Introduction}
Blazars are the most extreme objects in the family of active galactic nuclei (AGNs). With their jets closely aligned to our line of sight, they represent the best target for the study of both the physics of particle acceleration and the role played by magnetic fields in these extreme plasma environments \citep[e.g.][]{Blandford1979, Marscher2008}.
Blazars include both flat spectrum radio quasars (FSRQs - with an optical spectrum dominated by strong emission lines) and BL Lac objects (BL Lacs - showing almost featureless spectra with occasional weak optical emission lines, \citet{Stickel1991}).
Blazar emission spans the entire electromagnetic spectrum from radio up to TeV $\gamma$-ray energies, and is mostly non-thermal in nature and linearly polarised, providing us with important insights into the magnetic field structure \citep[e.g.][]{Gabuzda1994, Gomez2008, Orienti2011, Hovatta2012, Casadio2019}.
 Blazars tend to show erratic variability across the electromagnetic spectrum on different timescales, ranging from years \citep{Ulrich1997} down to a few minutes \citep[e.g.][]{Aharonian2007, Albert2007b}.
A small subset of blazar sources exhibit possible quasi-periodic variability across the radio, optical, and X-ray emission bands \citep[e.g.][]{Villata2004, Hovatta2008, Valtonen2008, Wiita2011, King2013}.
Over the last few years, thanks to the \fermi\ Large Area Telescope (\fermi -LAT) continuous monitoring of the sky in the MeV-GeV energy range \citep{Atwood2009}, quasi-periodic variability has been investigated at $\gamma$-ray energies in a number of blazars \citep[e.g.][]{Prokhorov2017}. 
Such high-energy periodicity might be related to jet precession and/or modulation of the accretion rate onto the central engine(s). As such, the study of quasi-periodic variability at $\gamma$-ray energies can help shed light on fundamental issues such as the disc--jet connection and the nature of the jet's magnetic fields, and could provide further insight into gravitational wave production in binary super massive black hole (SMBH) systems \citep{Abbott2016}. 

In this context, the BL Lac object PG\,1553+113 has exhibited complex high-energy variability and has been detected at MeV/GeV $\gamma$-ray energies by the \fermi -LAT at a significance level above 10 $\sigma$ \citep{Abdo2009}. PG\,1553+113 is the first blazar for which a quasi-periodic pattern with a period of $\sim2.18\pm 0.08$ years \citep{Ackermann2015} has been observed in its $\gamma$-ray light curve, providing us with a unique opportunity to investigate this peculiar behaviour. 

PG\,1553+113 exhibits an almost featureless optical spectrum \citep{Miller1983, Falomo1990} and the emission at all wavelengths, from radio to very high energies (VHE; E$>$ 100\,GeV), can be attributed to the non-thermal jet emission, without significant contributions from other substructures (e.g.\, the disc, the corona, or the broad line region). The host galaxy is yet undetected and the redshift determination is still uncertain. 
Using a method based on Bayesian statistics and the extragalactic background light absorption effects on the VHE spectrum, \citet{Abramowski2015} obtained a value of $z=0.49\pm 0.04$. Additional indirect methods, such as the Ly$\alpha$ forest method and the absence of a break in the VHE spectrum, constrain the redshift to the range 0.3--0.6 \citep[e.g.][]{Qin2018, Landoni2014, Danforth2010, Prandini2010, Mazin2007}.
This source is classified as a high synchrotron peaked (HSP) blazar \citep{Giommi1995} and was detected by the HESS and MAGIC $\gamma$-ray Cherenkov telescopes in 2005 \citep{Aharonian2006, Albert2007a}.

The quasi-periodic trend found in the $\gamma$-ray light
curve of PG\,1553+113, although still debated  \citep[e.g.][]{Covino2019}, is matched by similar behaviour at lower frequencies. In addition to the periodicity in the $\gamma$-ray emission, hints of periodicities have been found in the optical emission light
curve spanning $\sim$10 years (2004-2014) obtained within the Tuorla optical monitoring program\footnote{\url{http://users.utu.fi/kani/1m/}} and in the 0.3--10\,keV light
curve obtained by XRT on board the \textit{Neil Gehrles Swift} satellite, although at a lower significance level due to the sparse observations \citep{Tavani2018}.

To investigate the radio properties of the innermost regions of the jet, where the $\gamma$ rays are thought to be produced, high-resolution very long baseline interferometry (VLBI) observations are needed. 
On VLBI scales, the source shows a compact nuclear region with jet structure in the northeast direction (position angle $\sim50^\circ$), that is well detected out to roughly 20 pc from the core region \citep{Rector2003}. Beyond this distance the jet becomes very faint and diffuse.
Although several multi-wavelength (MWL) observing campaigns of PG\,1553+113 \citep[e.g.][]{Raiteri2015, Hovatta2016, Itoh2016, Raiteri2017} were conducted in recent years, only sparse VLBI observations of PG\,1553+113 can be found in the literature, including the 15\,GHz observations within the MOJAVE database\footnote{\url{https://www.physics.purdue.edu/MOJAVE/}} and the TeV Blazars very long baseline array (VLBA) monitoring program led by B.~G.~Piner and P.~G.~Edwards\footnote{\url{http://whittierblazars.com/}}.
For these reasons, we conducted a systematic multi-frequency (15, 24, and 43\,GHz) VLBA monitoring of PG\,1553+113 covering a period between two consecutive maxima in the $\gamma$-ray activity during 2015--2017. 
The main purpose of this observing campaign is to investigate and characterise the parsec-scale source properties and their evolution with time as well as the possible connections with the observed flux density modulation.

In this paper we describe the observing campaign and the data analysis in Sect.~\ref{sec_observations}, while the main results are presented in Sect.~\ref{sec_results}. A general discussion and summary of our findings are presented in Sects.~\ref{sec_discussion} and~\ref{sec_conclusion}, respectively.
Throughout the paper we use a $\Lambda$CDM cosmology with $h = 0.71$, $\Omega_m = 0.27$, and $\Omega_\Lambda=0.73$ \citep{Komatsu2011}.
The spectral index $\alpha$ is defined as $S_{\nu}\propto\nu^{-\alpha}$, and all angles are measured from north to east. 
At a redshift of $z = 0.49$ \citep{Abramowski2015} 1\,mas corresponds to $\sim5.2$ pc.

\begin{table}
\centering
\begin{tiny}
\caption{Details about the different VLBA observing sessions.}
\label{tab_observations}
\setlength{\tabcolsep}{4.3pt}
\begin{tabular}{lcccl}
\hline
\hline
Exp. code & Obs. date & MJD &  Refant & Missing/flagged stations \\
&yyyy/mm/dd&&& (obs. band in brackets) \\
\hline
BL214   &       2015/02/02  &   57055   &       PT      &       HK, NL (K, Q)      \\
-                &      2015/02/25  &   57078   &       PT      &       SC (Q)     \\
BL215   &       2015/09/11      &       57276   &       PT      &       SC (Q)     \\
-                       &       2015/10/17      &       57312   &       OV      &       -       \\
-                       &       2016/01/03      &       57390   &       OV      &       PT, BR, MK (K, Q), HK (Q)   \\
-                       &       2016/02/20      &       57438   &       PT      &       HK (Q), MK (K)     \\
-                       &       2016/04/05      &       57483   &       PT      &       -       \\
-                       &       2016/05/17      &       57525   &       PT      &       SC (Q)     \\
-                       &       2016/06/19      &       57558   &       OV      &       SC (Q)     \\
-                       &       2016/09/19      &       57650   &       OV      &       SC (Q), HK ()Q     \\
-                       &       2016/10/11      &       57672   &       OV      &       BR, MK, HK (Q)      \\
-                       &       2017/01/02      &       57755   &       OV      &       MK, FD (Q)  \\
-                       &  2017/03/23   &       57835   &       OV      &       MK (Q)     \\
-                       &       2017/06/12      &       57916   &       OV      &       SC (Q)     \\
\hline
\end{tabular} 
\tablefoot{ Station codes: BR - Brewster, FD - Fort Davis, HK - Hancock, KP - Kitt Peak, LA - Los Alamos, MK - Mauna Kea, NL - North Liberty, OV - Owens Valley, PT - Pie Town, SC - St. Croix.}
\end{tiny}
\end{table}


\begin{figure*}
\begin{center}
\includegraphics[bb=10 0 145 748, scale=0.8, clip]{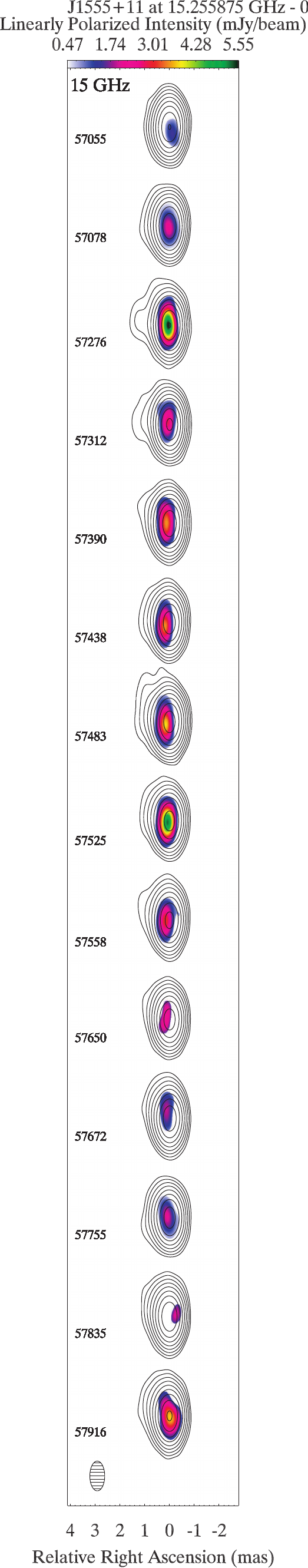}  
\includegraphics[bb=0 0 165 759, scale=0.8, clip]{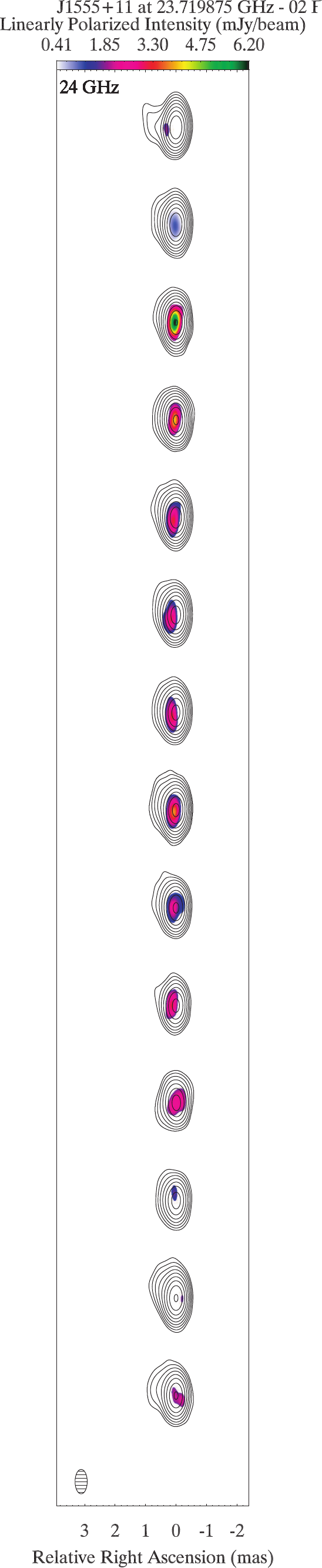}   
\includegraphics[bb=10 0 145 750, scale=0.8, clip]{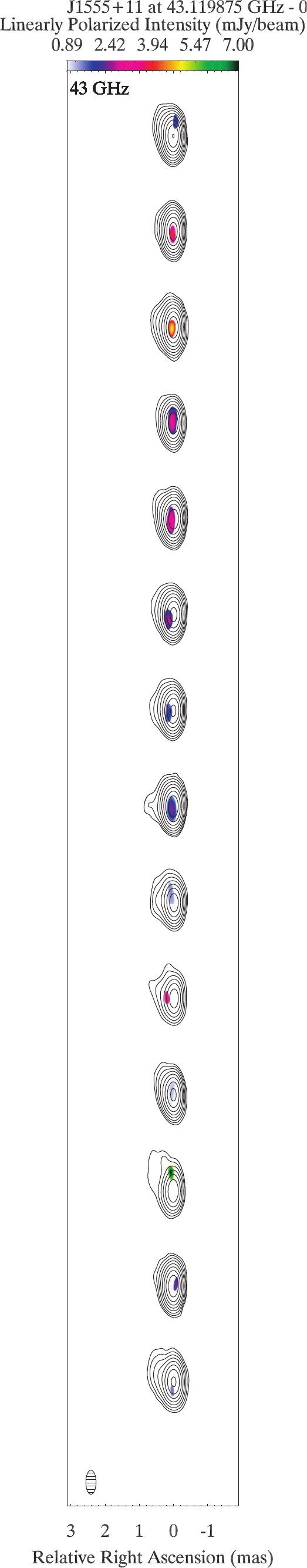}  
\end{center}
\caption{\small Natural weighted 15\,GHz (left panel), 24\,GHz (middle panel), and 43\,GHz (right panel) VLBA images of PG\,1553+113 from the first observing epoch (top image in each panel) to the fourteenth epoch (bottom image in each panel). The vertical separation between images is not proportional to the time difference between epochs. 
The images at each frequency have been convolved with a common beam with FWHM $0.6$\,mas $\times \, 1.2$\,mas, $0.4$\,mas $\times \, 0.8$\,mas and $0.3$\,mas $\times \, 0.7$\,mas at 15, 24 and 43\,GHz, respectively, as shown in the bottom-left corner in each panel. The overlaid lowest total intensity contour is at 0.4\%, 1.3\%, and 1.8\% of the peak at 15, 24, and 43\,GHz (see Table~\ref{table_data}), respectively, with the following contours a factor of two higher. The colour scale represents the linearly polarised intensity.}
\label{maps}
\end{figure*}

\begin{figure}
\includegraphics[width=1.0\columnwidth]{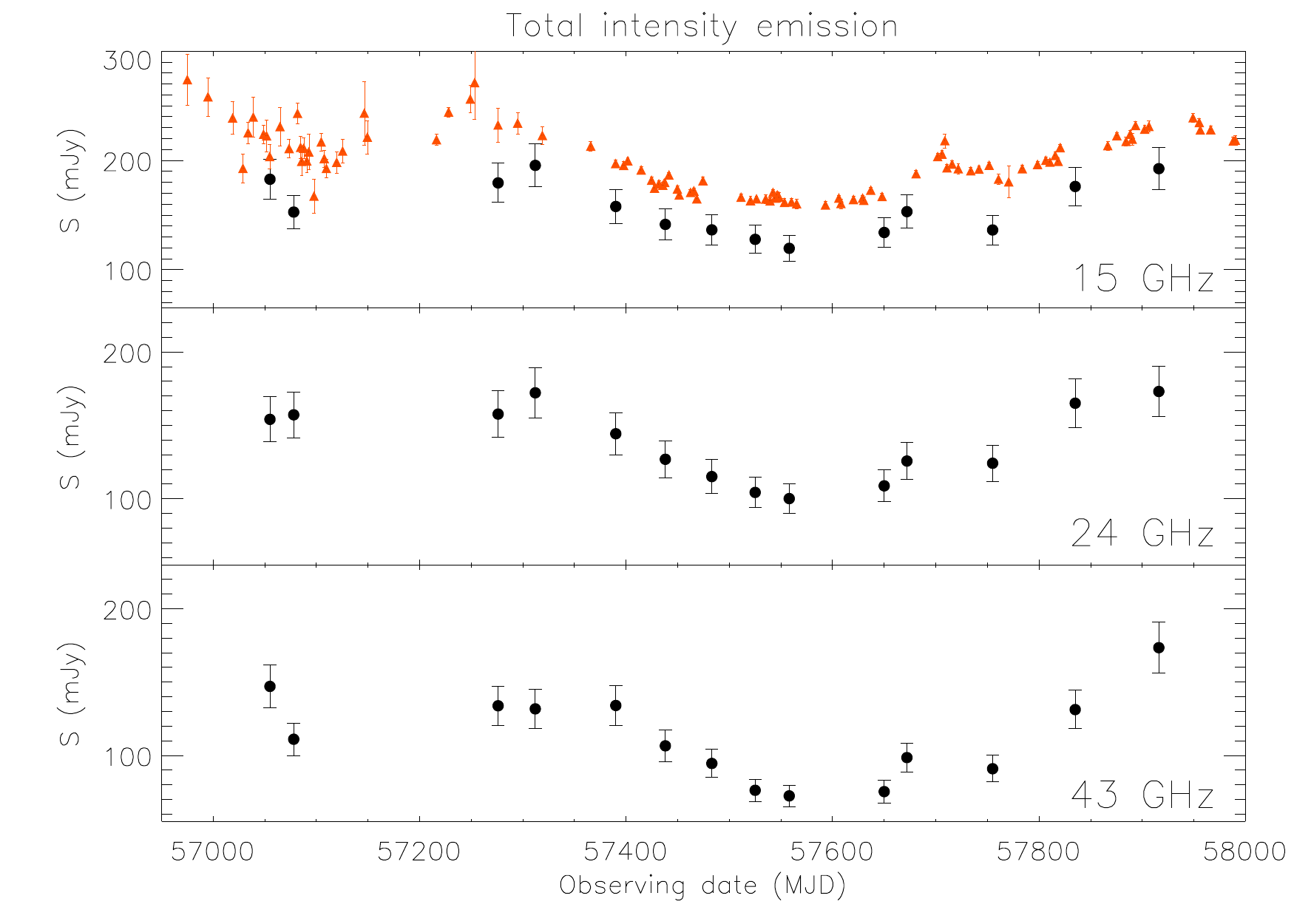}
   
\includegraphics[width=1.0\columnwidth]{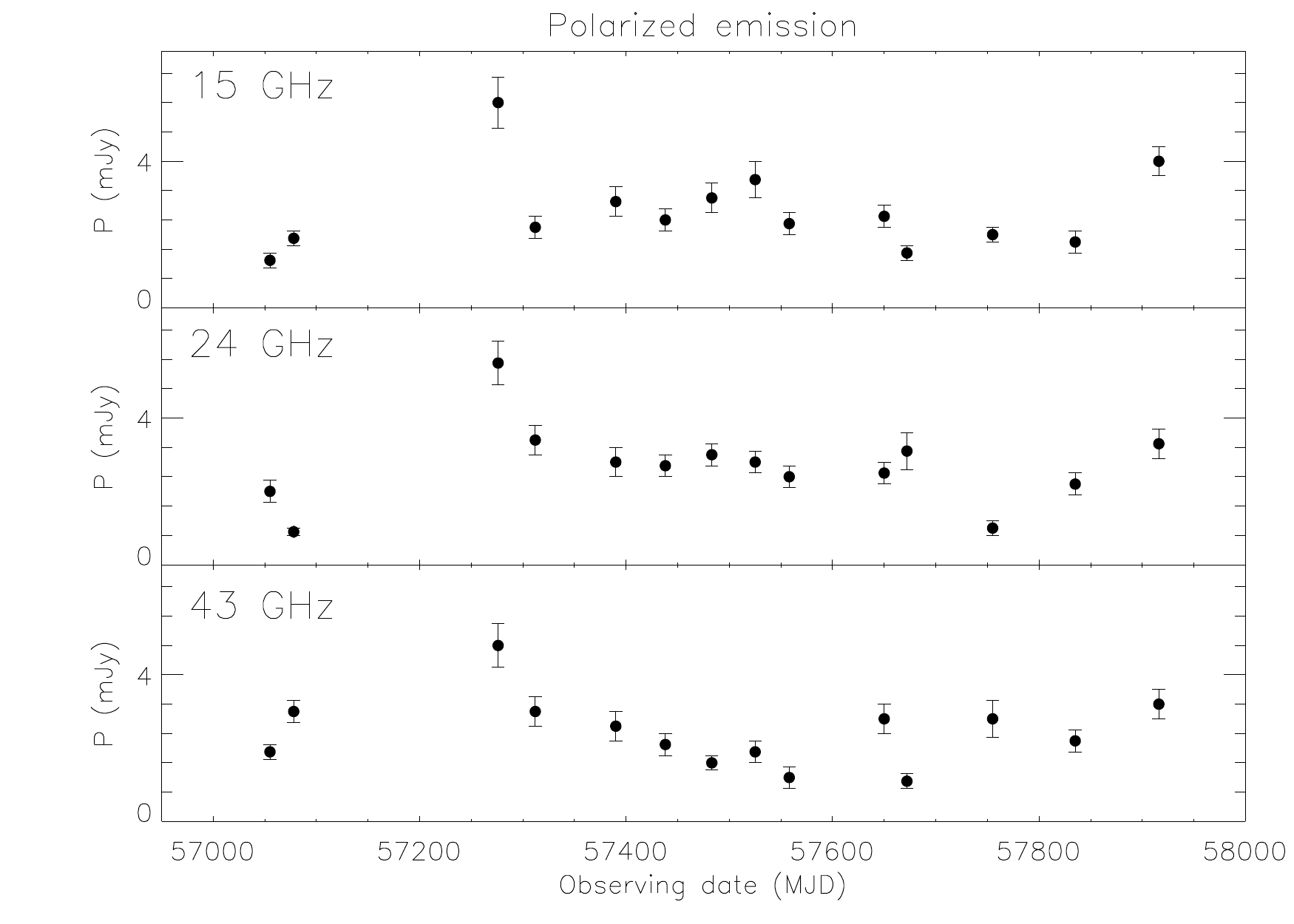}

\caption{Evolution with time of the total intensity (upper image) and polarised flux density (lower image) in the core region of PG\,1553+113 during the MJD 57055-57916 period. In each image we report the 15\,GHz (upper frame), 24\,GHz (middle frame), and 43\,GHz (lower frame) results. The overlaid red triangles in the upper image (top panel) represent the 15\,GHz OVRO light
curve in the MJD range 56974-57990.}
\label{light_curves}
\end{figure}

\begin{table*}
\begin{center}
\begin{tiny}
\caption{Summary of the total and polarised intensity parameters shown in Fig.~\ref{light_curves} and in the lower frame of Fig.~\ref{RM_plot}.}
\label{table_data}   
\setlength{\tabcolsep}{4.3pt}
\renewcommand{\arraystretch}{1.0}     
\begin{tabular}{lcccccccccccccccc}  
\hline\hline                 
Frequency & Obs. date & MJD & $S_{peak}$\tablefootmark{a} & $\sigma_{S_{peak}}$\tablefootmark{b} & $S_{core}$\tablefootmark{c} & $\sigma_{S_{core}}$\tablefootmark{d} & $S_{total}$\tablefootmark{e} & $\sigma_{S_{total}}$\tablefootmark{f} & P\tablefootmark{g} & $\sigma_P$\tablefootmark{h} & m\tablefootmark{i} & $\sigma_m$\tablefootmark{j} & $\chi$\tablefootmark{k} & $\sigma_{\chi}$\tablefootmark{l} & a\tablefootmark{m} & b/a\\
& yy/mm/dd &  & (mJy) & (mJy) & (mJy) & (mJy) & (mJy) & (mJy) & (mJy) & (mJy) & (\%) & (\%) & (deg) & (deg) & (mas) &  \\
\hline\hline     
&&&&&&&&&&&&&&\\
15\,GHz &       2015/02/02      &       57055   &       165.7   &       16.6    &       182.8   &       18.3    &       187.1   &       18.7    &       1.3     &       0.2     &       0.7     &       0.1     &       52.8    &       6.5     &       0.3     &       0.8     \\
&       2015/02/25      &       57078   &       140.0   &       14.0    &       152.8   &       15.3    &       159.9   &       16.0    &       1.9     &       0.2     &       1.3     &       0.2     &       22.6    &       5.1     &       0.2     &       0.9     \\
&       2015/09/11      &       57276   &       168.2   &       16.8    &       179.5   &       17.9    &       187.7   &       18.8    &       5.6     &       0.7     &       3.1     &       0.5     &       -       &       -       &       0.2     &       0.8     \\
&       2015/10/17      &       57312   &       182.8   &       18.3    &       195.5   &       19.6    &       203.2   &       20.3    &       2.2     &       0.3     &       1.1     &       0.2     &       -       &       -       &       0.2     &       0.9     \\
&       2016/01/03      &       57390   &       136.9   &       13.7    &       157.9   &       15.8    &       165.4   &       16.5    &       2.9     &       0.4     &       1.8     &       0.3     &       -       &       -       &       0.3     &       0.7     \\
&       2016/02/20      &       57438   &       121.9   &       12.2    &       141.5   &       14.1    &       149.2   &       14.9    &       2.4     &       0.3     &       1.7     &       0.2     &       -       &       -       &       0.3     &       0.8     \\
&       2016/04/05      &       57483   &       119.0   &       11.9    &       136.5   &       13.7    &       142.8   &       14.3    &       3.0     &       0.4     &       2.2     &       0.4     &       -       &       -       &       0.3     &       0.7     \\
&       2016/05/17      &       57525   &       111.9   &       11.2    &       127.8   &       12.8    &       136.7   &       13.7    &       3.5     &       0.5     &       2.7     &       0.5     &       -       &       -       &       0.3     &       0.7     \\
&       2016/06/19      &       57558   &       103.6   &       10.4    &       119.6   &       12.0    &       125.6   &       12.6    &       2.3     &       0.3     &       1.9     &       0.3     &       -       &       -       &       0.4     &       0.7     \\
&       2016/09/19      &       57650   &       120.5   &       12.1    &       134.1   &       13.4    &       138.5   &       13.9    &       2.5     &       0.3     &       1.9     &       0.3     &       154.9   &       5.6     &       0.3     &       0.9     \\
&       2016/10/11      &       57672   &       140.6   &       14.1    &       153.2   &       15.3    &       157.4   &       15.7    &       1.5     &       0.2     &       1.0     &       0.2     &       -       &       -       &       0.3     &       0.8     \\
&       2017/01/02      &       57755   &       123.9   &       12.4    &       136.4   &       13.6    &       140.4   &       14.0    &       2.0     &       0.2     &       1.5     &       0.2     &       69.1    &       6.6     &       0.4     &       0.6     \\
&       2017/03/23      &       57835   &       161.4   &       16.1    &       176.2   &       17.6    &       183.1   &       18.3    &       1.8     &       0.3     &       1.0     &       0.2     &       -       &       -       &       0.3     &       0.8     \\
&       2017/06/12      &       57916   &       174.6   &       17.5    &       192.5   &       19.3    &       201.2   &       20.1    &       4.0     &       0.4     &       2.1     &       0.3     &       -       &       -       &       0.3     &       0.7     \\
&&&&&&&&&&&&&&\\
24\,GHz &       2015/02/02      &       57055   &       133.8   &       13.4    &       154.1   &       15.4    &       160.4   &       16.0    &       2.0     &       0.3     &       1.3     &       0.2     &       49.5    &       6.9     &       0.2     &       0.6     \\
&       2015/02/25      &       57078   &       137.3   &       13.7    &       157.2   &       15.7    &       164.7   &       16.5    &       0.9     &       0.1     &       0.6     &       0.1     &       128.0   &       6.4     &       0.2     &       0.7     \\
&       2015/09/11      &       57276   &       142.5   &       14.3    &       157.7   &       15.8    &       164.5   &       16.5    &       5.5     &       0.6     &       3.5     &       0.5     &       -       &       -       &       0.2     &       0.7     \\
&       2015/10/17      &       57312   &       154.8   &       15.5    &       172.2   &       17.2    &       183.6   &       18.4    &       3.4     &       0.4     &       2.0     &       0.3     &       -       &       -       &       0.2     &       0.7     \\
&       2016/01/03      &       57390   &       117.8   &       11.8    &       144.3   &       14.4    &       152.0   &       15.2    &       2.8     &       0.4     &       1.9     &       0.3     &       -       &       -       &       0.3     &       0.7     \\
&       2016/02/20      &       57438   &       104.7   &       10.5    &       126.9   &       12.7    &       133.3   &       13.3    &       2.7     &       0.3     &       2.2     &       0.3     &       -       &       -       &       0.3     &       0.7     \\
&       2016/04/05      &       57483   &       94.4    &       9.4     &       115.1   &       11.5    &       122.3   &       12.2    &       3.0     &       0.3     &       2.6     &       0.4     &       -       &       -       &       0.3     &       0.6     \\
&       2016/05/17      &       57525   &       88.8    &       8.9     &       104.3   &       10.4    &       113.2   &       11.3    &       2.8     &       0.3     &       2.7     &       0.4     &       -       &       -       &       0.2     &       0.5     \\
&       2016/06/19      &       57558   &       85.9    &       8.6     &       100.1   &       10.0    &       102.3   &       10.2    &       2.4     &       0.3     &       2.4     &       0.4     &       -       &       -       &       0.2     &       0.8     \\
&       2016/09/19      &       57650   &       91.4    &       9.1     &       108.8   &       10.9    &       110.4   &       11.0    &       2.5     &       0.3     &       2.3     &       0.4     &       147.5   &       8.5     &       0.3     &       0.7     \\
&       2016/10/11      &       57672   &       104.9   &       10.5    &       125.7   &       12.6    &       129.4   &       12.9    &       3.1     &       0.5     &       2.5     &       0.4     &       -       &       -       &       0.3     &       0.7     \\
&       2017/01/02      &       57755   &       104.8   &       10.5    &       124.2   &       12.4    &       126.9   &       12.7    &       1.0     &       0.2     &       0.8     &       0.2     &       46.2    &       7.2     &       0.3     &       0.7     \\
&       2017/03/23      &       57835   &       144.1   &       14.4    &       165.1   &       16.5    &       177.0   &       17.7    &       2.2     &       0.3     &       1.3     &       0.2     &       -       &       -       &       0.2     &       0.7     \\
&       2017/06/12      &       57916   &       144.7   &       14.5    &       173.1   &       17.3    &       179.3   &       17.9    &       3.3     &       0.4     &       1.9     &       0.3     &       -       &       -       &       0.3     &       0.7     \\
&&&&&&&&&&&&&&\\
43\,GHz &       2015/02/02      &       57055   &       112.1   &       11.2    &       147.1   &       14.7    &       149.2   &       14.9    &       1.9     &       0.2     &       1.3     &       0.2     &       129.5   &       6.0     &       0.2     &       0.6     \\
&       2015/02/25      &       57078   &       96.8    &       9.7     &       111.1   &       11.1    &       113.7   &       11.4    &       3.0     &       0.3     &       2.7     &       0.4     &       3.7     &       6.1     &       0.2     &       0.6     \\
&       2015/09/11      &       57276   &       114.9   &       11.5    &       133.9   &       13.4    &       144.3   &       14.4    &       4.8     &       0.6     &       3.6     &       0.6     &       -       &       -       &       0.2     &       0.6     \\
&       2015/10/17      &       57312   &       109.6   &       11.0    &       131.8   &       13.2    &       137.7   &       13.8    &       3.0     &       0.4     &       2.2     &       0.4     &       -       &       -       &       0.1     &       0.8     \\
&       2016/01/03      &       57390   &       105.1   &       10.5    &       134.1   &       13.4    &       135.9   &       13.6    &       2.6     &       0.4     &       2.0     &       0.3     &       -       &       -       &       0.2     &       0.8     \\
&       2016/02/20      &       57438   &       86.4    &       8.6     &       106.6   &       10.7    &       115.5   &       11.5    &       2.1     &       0.3     &       2.0     &       0.3     &       -       &       -       &       0.2     &       0.7     \\
&       2016/04/05      &       57483   &       75.5    &       7.5     &       94.6    &       9.5     &       103.8   &       10.4    &       1.6     &       0.2     &       1.7     &       0.3     &       -       &       -       &       0.2     &       0.5     \\
&       2016/05/17      &       57525   &       65.1    &       6.5     &       76.3    &       7.6     &       82.1    &       8.2     &       1.9     &       0.3     &       2.5     &       0.4     &       -       &       -       &       0.2     &       0.5     \\
&       2016/06/19      &       57558   &       63.8    &       6.4     &       72.5    &       7.3     &       76.2    &       7.6     &       1.2     &       0.3     &       1.6     &       0.4     &       -       &       -       &       0.2     &       0.9     \\
&       2016/09/19      &       57650   &       64.9    &       6.5     &       75.4    &       7.6     &       81.3    &       8.1     &       2.8     &       0.4     &       3.7     &       0.6     &       168.8   &       6.7     &       0.3     &       0.6     \\
&       2016/10/11      &       57672   &       86.0    &       8.6     &       98.6    &       9.9     &       100.9   &       10.1    &       1.1     &       0.2     &       1.1     &       0.3     &       -       &       -       &       0.3     &       0.4     \\
&       2017/01/02      &       57755   &       69.1    &       6.9     &       91.0    &       9.1     &       98.8    &       9.9     &       2.8     &       0.5     &       3.1     &       0.6     &       99.2    &       7.1     &       0.3     &       0.5     \\
&       2017/03/23      &       57835   &       106.0   &       10.6    &       131.3   &       13.1    &       137.0   &       13.7    &       2.2     &       0.3     &       1.7     &       0.3     &       -       &       -       &       0.2     &       0.7     \\
&       2017/06/12      &       57916   &       134.3   &       13.4    &       173.5   &       17.4    &       188.2   &       18.8    &       3.2     &       0.4     &       1.9     &       0.3     &       -       &       -       &       0.2     &       0.8     \\
\hline                                  
\end{tabular}
\end{tiny}
\tablefoot{
\begin{tiny}
\newline
\tablefoottext{a}{Peak flux density ($S_{peak}$) in mJy;} \tablefoottext{b}{uncertainties on $S_{peak}$;}
\tablefoottext{c}{core region flux density ($S_{core}$) in mJy;} \tablefoottext{d}{uncertainties on $S_{core}$;}
\tablefoottext{e}{Total flux density ($S_{total}$) in mJy;} \tablefoottext{f}{uncertainties on $S_{total}$;}
\tablefoottext{g}{polarised flux density (P) in mJy;} \tablefoottext{h}{uncertainties on P;}
\tablefoottext{i}{fractional polarisation (m);} \tablefoottext{j}{uncertainties on m;}
\tablefoottext{k}{EVPAs (deg);} \tablefoottext{l}{uncertainties on the EVPAs;}
\tablefoottext{m}{a and b are the FWHM of the major and minor axes of the modelfit Gaussian components (mas).}
\end{tiny}
}
\end{center}
\end{table*}

\section{Observations and data analysis} 
\label{sec_observations}
We monitored PG\,1553+113 with the VLBA at 15, 24, and 43\,GHz in full polarisation from February 2015 to June 2017. The observations are based on two related observing projects: two six-hour observing sessions during the period of its maximum $\gamma$-ray activity in 2015 (BL214) and a full two-year monitoring program with roughly bimonthly runs (BL215, 54-h in total). 
Table~\ref{tab_observations} reports the log of the observations. In some epochs one or more VLBA stations were missing or flagged due to technical problems (as reported in the last column in Table~\ref{tab_observations}); when a station is missing only for a specific observing band, it is indicated in brackets.
We were able to obtain absolute electric vector position angle (EVPA) orientations for the four observing epochs (MJD 57055, 57078, 57650, 57755) for which we had quasi-simultaneous Karl G.\,Jansky very large array (JVLA) observations at similar frequencies. We used the source J1310+3220 as the instrumental polarisation calibrator.
We also made use of the 15\,GHz observations provided by the Owens Valley Radio Observatory (OVRO) blazar monitoring program \citep{Richards2011}. PG\,1553+113 has been monitored with OVRO since 2008 with a cadence of about two observations per week. In this work we use the observations over the MJD range 54696-58153\footnote{\url{http://www.astro.caltech.edu/ovroblazars/}}.

We used the software package Astronomical Image Processing System \citep[AIPS; ][]{Greisen2003} for the data calibration, the fringe-fitting, and the detection of cross-polarised fringes. We determined the instrumental polarisation leakages (D-terms) with the AIPS task LPCAL.
For the final images we used the CLEAN and self-calibration procedures in the \texttt{DIFMAP} software package  \citep{Shepherd1997}. The core flux densities were obtained by fitting elliptical Gaussian components to the calibrated visibilities for each epoch at each frequency in \texttt{DIFMAP}. 
In this work the core is identified as a bright and stationary feature at the upper end of the jet, with a roughly flat spectrum and high brightness temperature (of the order of $10^{10}$\,K, see Sect.~\ref{sec_Tb}).
\\
For producing the polarisation images presented in this work we made use of IDL routines developed by J.~L.~G\'omez and for the ridge-line analysis the HEADACHE\footnote{\url{https://github.com/junliu/headache}} python package developed by J.~Liu was used.

\begin{figure*}
\begin{center}
\includegraphics[bb=0 5 353 252, scale=0.85, clip]{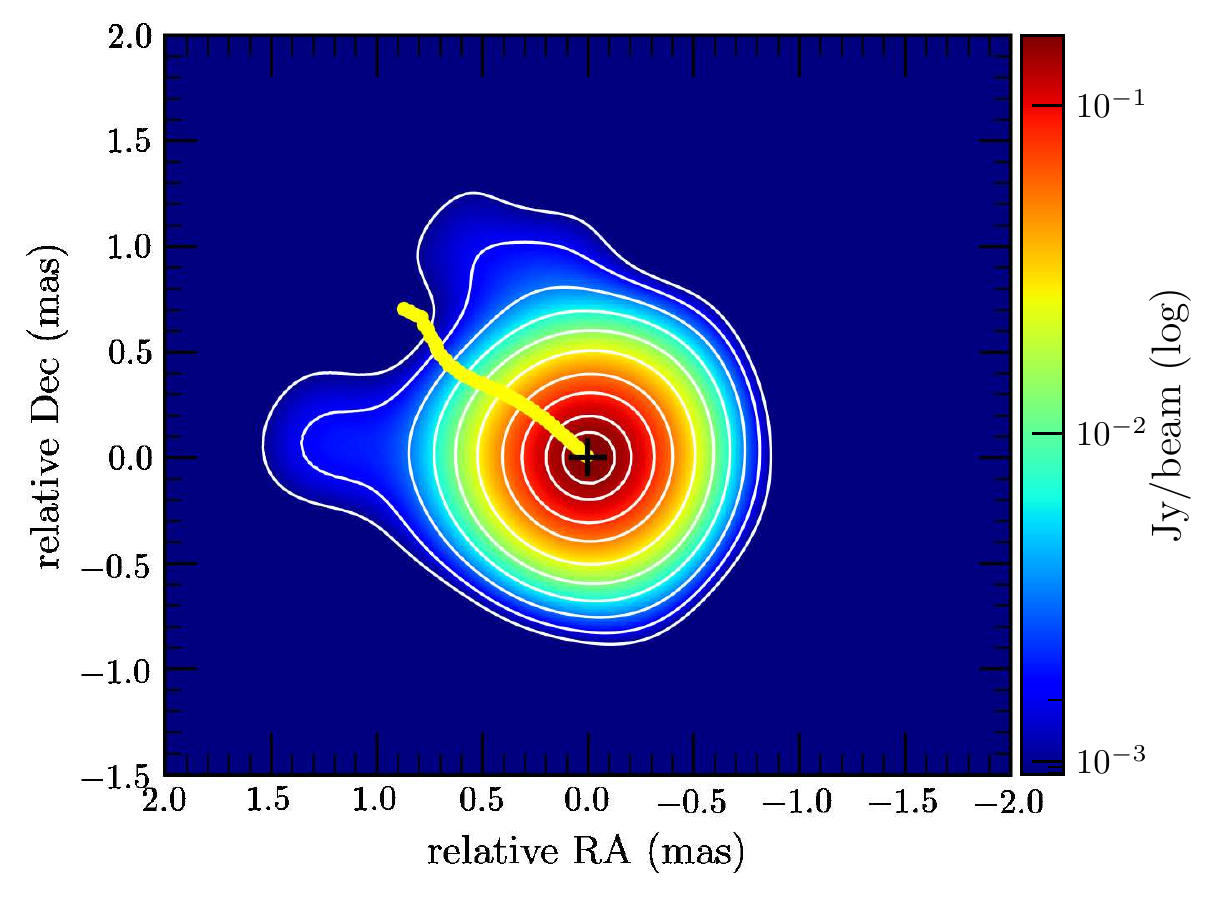}  
\includegraphics[bb=0 5 348 343, scale=0.6, clip]{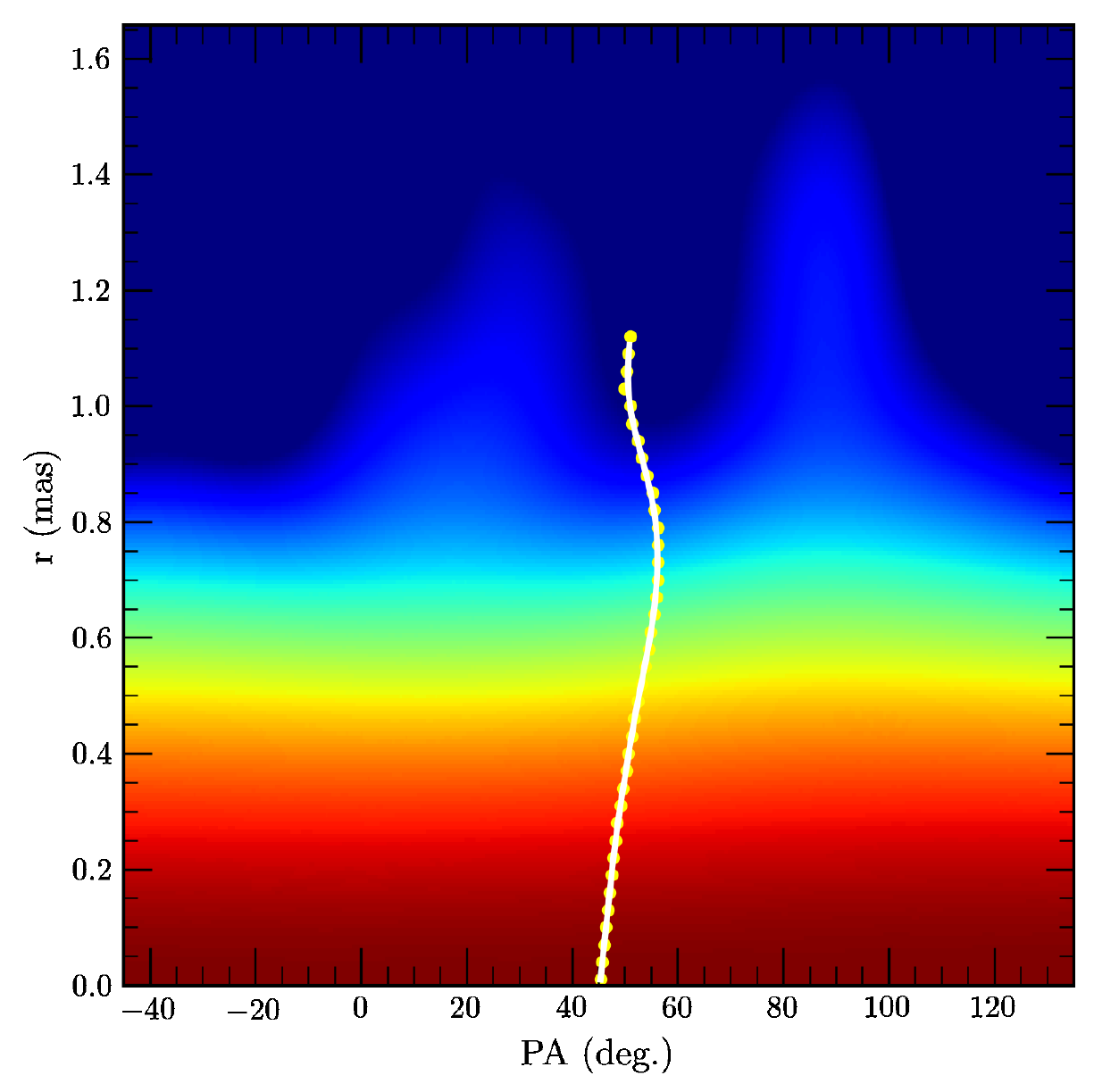}
\end{center}
\caption{\small Left image: 15\,GHz total intensity (contours and colour scale) image of PG\,1553+113 during the third observing epoch (MJD 57276), convolved with a $0.6$\,mas $\times \, 0.6$\,mas circular beam. The overlaid lowest total intensity contour is at 0.6\% of the peak, with the following contours a factor of two higher. The ridge line is overlaid as a yellow line. Right image: Same 15\,GHz image in polar coordinates. The colour scale represents the total intensity emission and the white line represents the ridge line.}
\label{ridge_3rd_epoch}
\end{figure*}

\begin{figure}
\includegraphics[width=1.0\columnwidth]{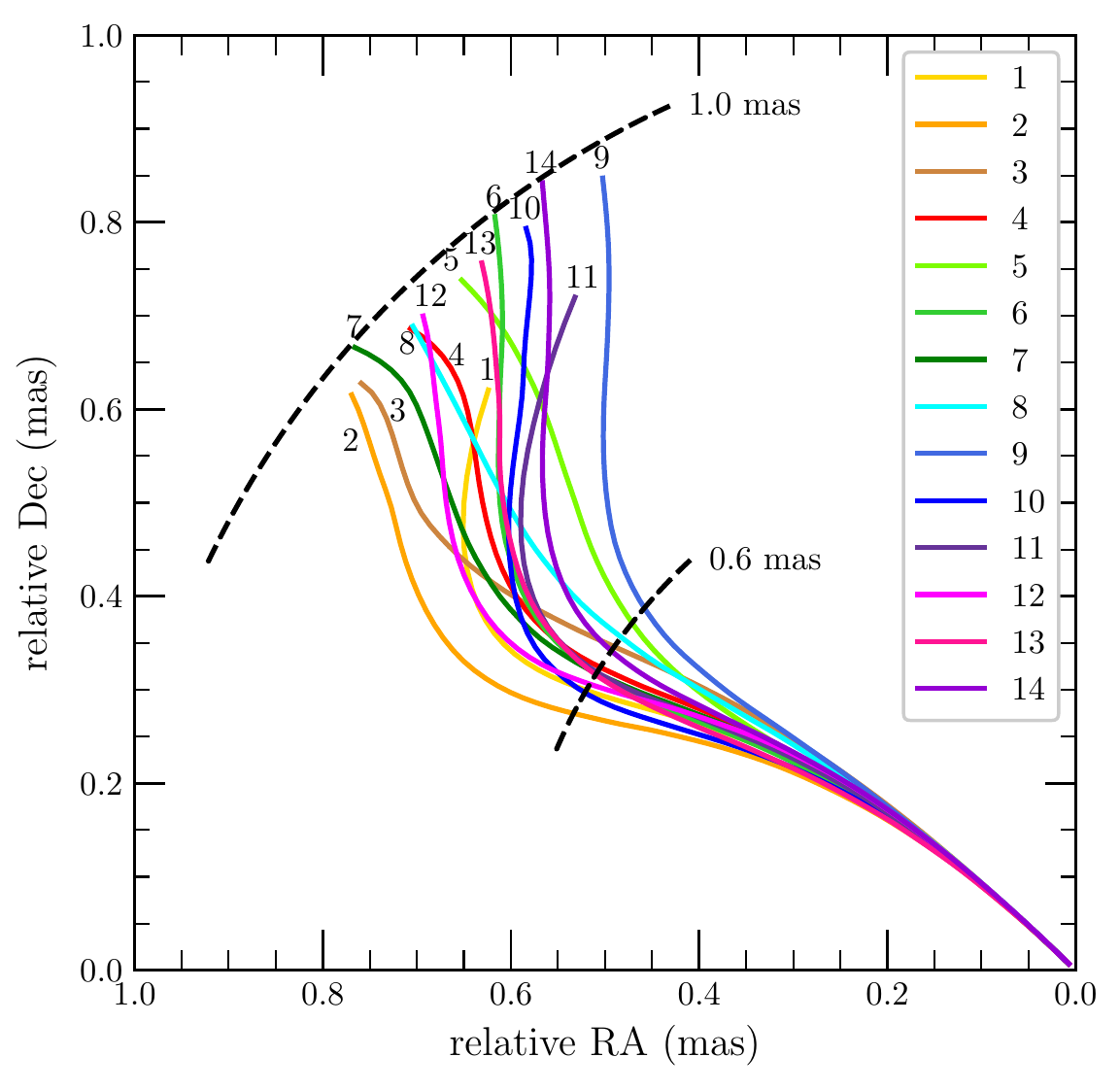}
    
\includegraphics[width=1.0\columnwidth]{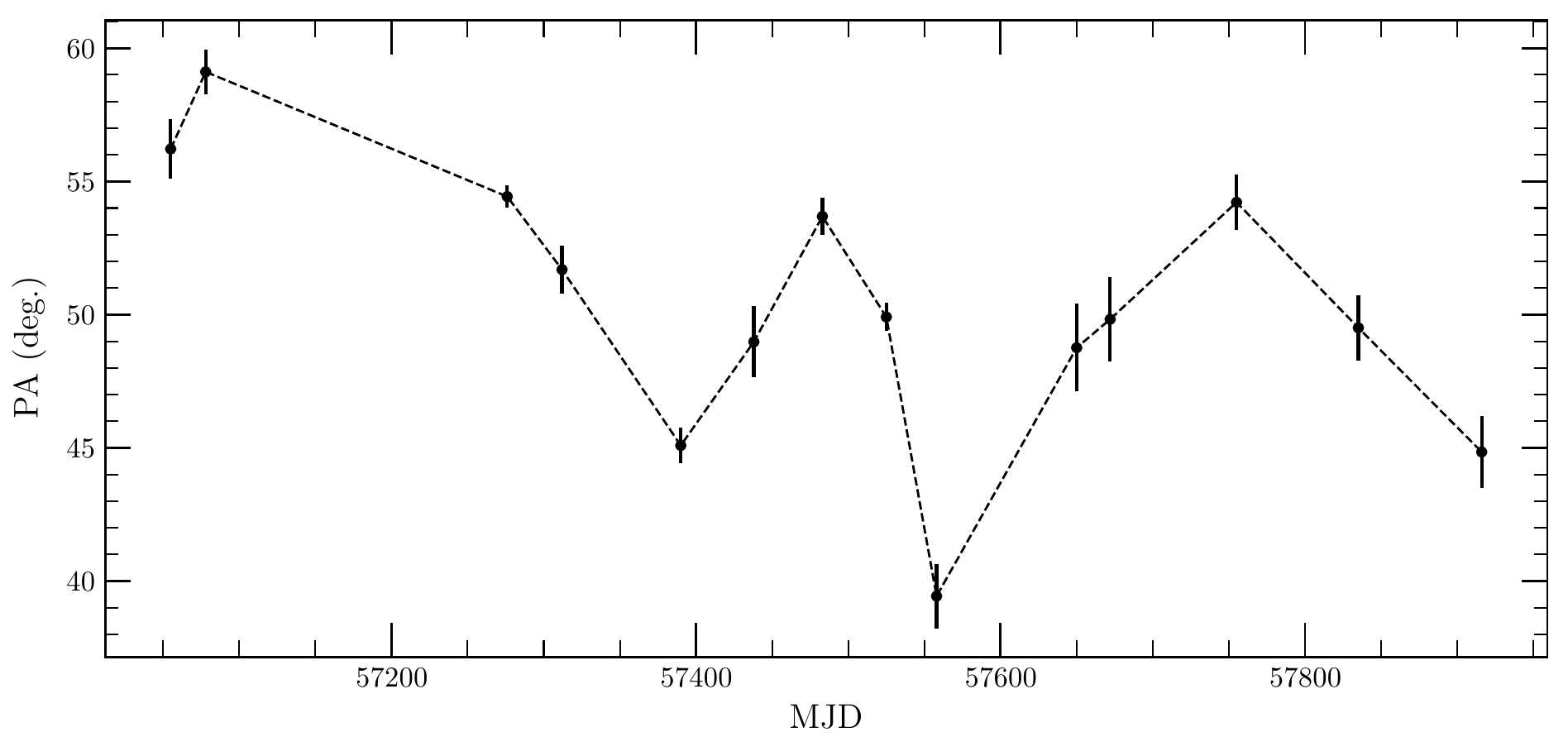}

\caption{Upper image: Ridge lines of PG\,1553+113 for all 14 observing epochs at 15\,GHz. Each epoch is represented by a different colour and is labelled with a progressive number. The two dashed arcs represent 0.6\,mas and 1\,mas from the core region. Lower image: Jet direction across the observing epochs represented by the average ridge-line angle (see Sect.~\ref{sec_ridgeline} for more details).}
\label{ridgelines}
\end{figure}

\section{Results} 
\label{sec_results}

\subsection{Images}
In Fig.~\ref{maps} we show the natural weighted images for PG\,1553+113 obtained during our 2015--2017 VLBA observing campaign at 15\,GHz (left panel), 24\,GHz (middle panel), and 43\,GHz (right panel). The source shows a compact core region with a northeast jet structure detected up to $\sim1.5$\,mas ($\sim7.8$\,pc in linear size) at 15\,GHz and $\sim1$\,mas ($\sim5.2$\,pc) at 43\,GHz. 
This morphology is in agreement with 15\,GHz MOJAVE images, as well as with the 22 and 43\,GHz results by \citet{Piner2018}. 
The jet position angle (PA) has an average value of $\sim50^{\circ}$. Further details and constrains about the PA across the different observing epochs are provided in Sect.~\ref{sec_ridgeline}.
Colours represent the linear polarised emission, which is mainly detected in the core region, while the black contours represent the total intensity emission. 

\subsection{Light curves}
In Fig.~\ref{light_curves} we show the core region total (upper image) and polarised (lower image) intensity light
curves.
In each image results at 15, 24, and 43\,GHz are reported in single frames from top to the bottom, respectively. 

We monitored PG\,1553+113 between February 2015 to June 2017 with a cadence of about 2 months, and with an approximately six-month gap between the second and third observing runs. Flux densities are reported in Table~\ref{table_data}.
We find two maxima in the total intensity emission (upper image in Fig.~\ref{light_curves}) around MJD 57312 (Oct 2015) and MJD 57916 (June 2017), that is, at the beginning and the end of the observing period, and a minimum around 57558 (Jun 2016).
The total intensity light curves are found to vary in the range 120-200 mJy at 15\,GHz, 100-170 mJy at 24\,GHz, and 70-170 mJy at 43\,GHz.
The variability is higher at 43\,GHz, where the fractional variability amplitude \citep[$F_{var}$, ][]{Vaughan2003} is $0.26 \pm  0.03$, while at 24 and 15\,GHz $F_{var}$ is $0.16 \pm 0.03$ and $0.13 \pm 0.03$, respectively.
Red triangles in the first frame in the top image of Fig.~\ref{light_curves} represent the 15\,GHz OVRO light
curve.

As shown in the lower image in Fig.~\ref{light_curves} we detect polarised emission from the source core region with an average value of $2.6$, $2.7$, and $2.5$ mJy at 15, 24, and 43\,GHz, respectively.
The polarised emission is also found to be variable, with $F_{var} \simeq 0.40\pm0.05$ at all three observing frequencies. During MJD 57276 a peak in the polarised emission of $5.6\pm0.7$, $5.5\pm0.6,$ and $4.8\pm0.6$\,mJy is detected at 15, 24, and 43\,GHz, respectively. 
The fractional polarisation varies in the range $0.7-3.1\%$ at 15\,GHz, $0.6-3.5\%$ at 24\,GHz, and $1.1-3.7\%$ at 43\,GHz. The average fractional polarisation is 1.7\%, 2.0\%, and 2.2\% at 15, 24, and 43\,GHz, respectively.

To asses the significance of the variability observed both in the total and polarised intensity light
curves, we performed a $\chi^2$ analysis by testing the null hypothesis of non-variability (with the flux density being constant around the average value). In all cases the null hypothesis can be rejected with a confidence level above $3\sigma$.

\subsection{Jet total intensity ridge line}
\label{sec_ridgeline}
It is apparent upon visual inspection of the source images (Fig.~\ref{maps}) that the jet direction varies across epochs. Given that the extended emission of the source on VLBI scales is not prominent and well-defined, it is not straightforward to represent the jet brightness distribution by means of Gaussian model-fit components.
In order to constrain the jet direction and to further quantify its variation with time, we calculate the total intensity ridge line in each epoch at 15\,GHz after convolving the images with a circular beam with a $0.6$\,mas radius. 
We use the method proposed by \citet{Pushkarev2017} by taking azimuthal slices along the jet in a polar coordinate system centred on the core. We look for the half-point  that divides the intensity integrated along the arc into two equal sections.
The slices along the jet direction are taken in the image plane in stpdf of $0.03$\,mas and only pixels with a signal-to-noise ratio of S/N $>10$ are taken into account.
This approach is particularly suitable in the case of a jet showing a limb-brightened structure in total intensity, that is a distinctive feature of HSP blazars \citep[e.g.][]{Giroletti2008, Piner2010, Lico2014}. As an example, in the left panel of Fig.~\ref{ridge_3rd_epoch} we present the 15\,GHz total intensity image obtained during MJD 57276, convolved with a $0.6$\,mas $\times$ $0.6$\,mas circular beam, with the ridge line (yellow line) overlaid. The limb brightened structure is clearly visible. The same image is shown in polar coordinates in the right panel of Fig.~\ref{ridge_3rd_epoch}, where the total intensity limb-brightened structure is also clearly visible (blue colour).

In the upper panel of Fig.~\ref{ridgelines} we present the ridge lines obtained for all of the 14 observing epochs in a single plot with different colours, where the jet wandering can clearly be seen. Overall, the ridge line extends from 0.9 to $\sim1$ mas from the core region.  
The jet direction in each epoch is determined by averaging all the angles of each ridge-line point between 0.6\,mas from the core region (i.e. beyond the beam radius) and the outermost ridge-line point ($\lesssim1$\,mas). All angles thus obtained are shown in the lower panel of Fig.~\ref{ridgelines} and reported in Table~\ref{tab_PA_SI}.
In principle we could also extract information about the ridge-line curvature. However, given that the jet of PG\,1553+113 between 15 and 43\,GHz is not particularly extended and is not well defined (core-dominated source), any estimates regarding the curvature would be largely uncertain.

To estimate the  angle of the funnel in which the jet wobbles, we first align the 15\,GHz images (convolved with a common beam with FWHM $0.6$\,mas $\times \, 1.2$\,mas) according to the position of the fitted core component, and after subtracting the core component emission we stack all of the residual images. Only those pixels with a S/N $>8$ are taken into account.  
In the residual stacked image (Fig.~\ref{residuals}), the jet-emitting region extends up to $\sim1$\,mas ($\sim5.2$\,pc) in a funnel with an angle of $\phi \sim 100$\,deg. 
A total intensity limb-brightened structure is clearly visible, with the southeast limb being brighter (peak flux density $\sim3.7$ mJy/beam) than the northern one (peak flux density $\sim2.5$ mJy/beam).

\subsection{Intrinsic polarisation angle and rotation measure}
\label{polarization_sec}
\begin{figure}
\includegraphics[bb=18 8 485 340, width=1.0\columnwidth, clip]{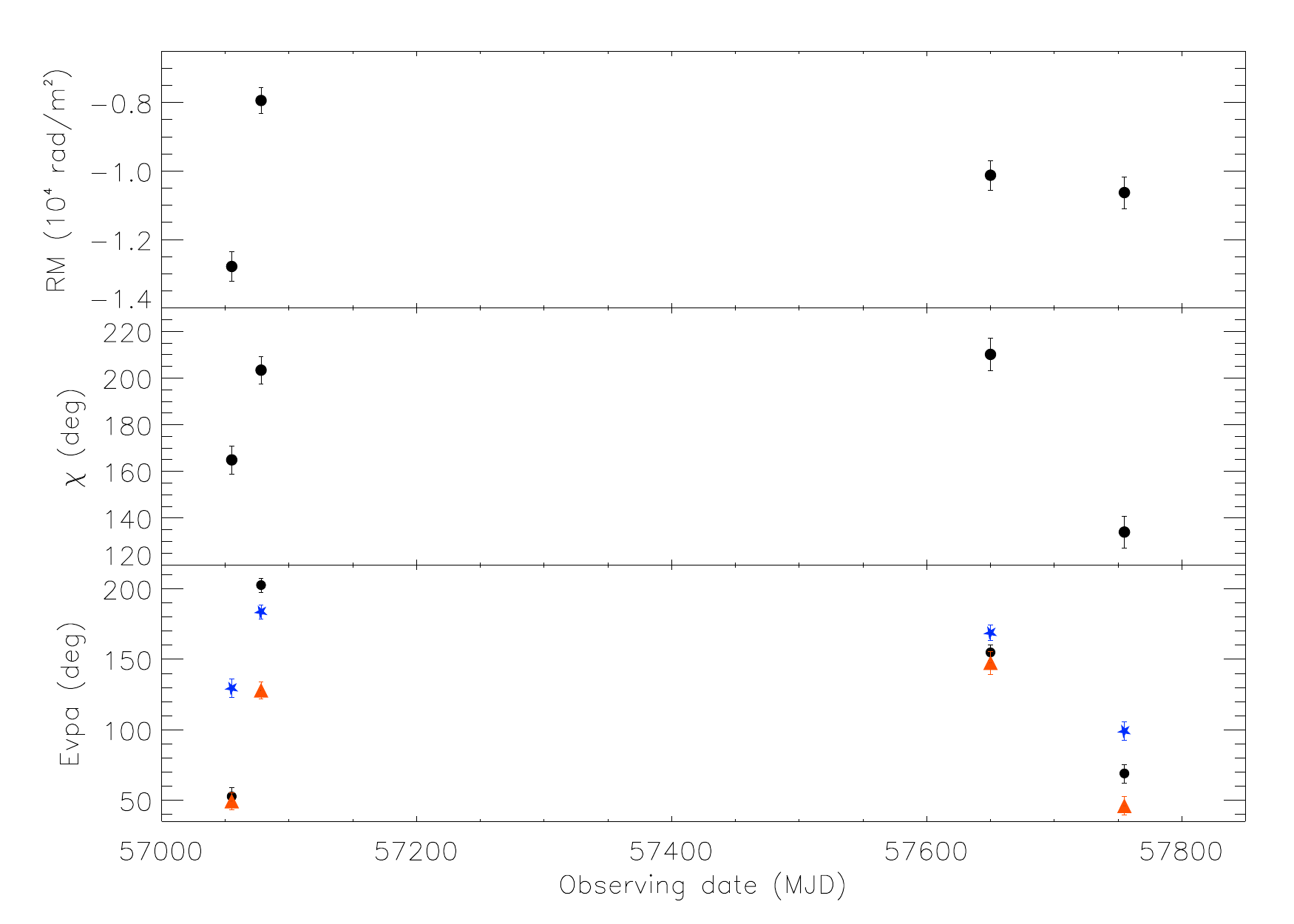}
    
\caption{Core region RM (upper panel), intrinsic polarisation angle (middle panel), and absolute EVPA orientation (lower panel) values during the epochs MJD 57055, 57078, 57650, and 57755. In the lower panel 15, 24, and 43\,GHz are represented by black circles, red triangles, and blue star symbols, respectively.} 
\label{RM_plot}
\end{figure}

Due to the lack of polarisation calibrators on VLBI scales, in order to obtain the absolute orientation of the EVPAs, a comparison with quasi-simultaneous single-dish or JVLA observations is required. The final EVPA values for the four epochs for which close in time JVLA observations were available (see Sect.~\ref{sec_observations}) are shown in the bottom panel of Fig.~\ref{RM_plot} for 15 (black circles), 24 (red triangles), and 43 (blue stars) GHz. The EVPAs are found to be variable across the different observing epochs, and change from being roughly aligned ($\sim50^{\circ}$) to approximately transverse ($\sim150^{\circ}$) to the jet axis. 

Because of the effect of Faraday rotation, which occurs when a polarised wave propagates through a magnetised plasma, the observed polarisation angle ($\chi_{obs}$), at a given observing wavelength $\lambda$ is rotated with respect to the intrinsic angle ($\chi_{int}$). 
Taking this effect into account is essential for obtaining the intrinsic polarisation angle as well as the intrinsic magnetic field orientation.
For a Faraday rotating medium external to the emitting region $\chi_\mathrm{obs} = \chi_\mathrm{int} + RM\times \, \lambda^2$, with RM being the rotation measure defined as: 
\begin{equation}
RM = 812 \int n_e \,\, \textbf{B}_{\parallel} \cdot dl \ \ \ [\mbox{rad \ m}^{-2}]
,\end{equation}
where $n_e$ is the electron density (cm$^{-3}$), $\textbf{B}_{\parallel}$ is the component of the magnetic field parallel to our line of sight (mG), and $dl$ the path length (pc). 
It is evident, within this theoretical framework, that a linear dependence exists between $\chi_{obs}$ and $\lambda^{2}$. We therefore compute linear fits to our observed EVPAs at 15, 24, and 43\,GHz to obtain estimates of $\chi_{int}$ and RM.

The values of RM and $\chi_{int}$ obtained for these four observing epochs are presented in Fig.~\ref{RM_plot} (top and middle panels, respectively). 
The average RM value is $\sim-1.0 \pm 0.4 \times 10^4$ rad \ m$^{-2}$, varying between $-1.3$ and $-0.8\times \, 10^4$ rad \ m$^{-2}$. 
The intrinsic polarisation angle is also variable across the different observing epochs, varying between $135^\circ \pm 7^\circ$ (i.e. roughly transverse to the jet axis) and $210^\circ \pm 7^\circ$ (i.e. roughly parallel to the jet axis).

\subsection{Brightness temperature measurements}
\label{sec_Tb}
By fitting the core brightness distribution in the uv-plane with an elliptical Gaussian component, via the modelfit procedure in \texttt{DIFMAP}, we can determine the observed rest-frame core brightness temperature $T_{\rm B,vlbi}^{\rm obs}$ \citep[e.g.][]{Tingay1998}: 
\begin{equation}
T_{\rm B,vlbi}^{\rm obs}=1.22 \times \, 10^{12} \frac{S_{core}(1+z)}{\theta_{maj}\theta_{min}\nu^2} \ \ \ [\mbox{K}]
,\end{equation}
where $S_{core}$ corresponds to the fitted core flux density (Jy) for a given observing frequency $\nu$ (GHz), $\theta_{maj}$ and $\theta_{min}$ are the FWHM (mas) of the major and minor axes of the elliptical Gaussian core component, and $z$ is the redshift.
The resulting $T_{\rm B,vlbi}^{\rm obs}$ values are reported in Table~\ref{tab_T_B}, where $T_{\rm B,vlbi}^{\rm obs}(\rm max)$, $T_{\rm B,vlbi}^{\rm obs}(\rm min),$ and $T_{\rm B,vlbi}^{\rm obs}(\rm avg)$ represent the maximum, the minimum, and average values at each frequency across the different observing epochs.

We also estimate the rest-frame core variability brightness temperature $T_{\rm B,var}^{\rm obs}$  using a variability timescale of $\tau\sim300$ days, which is the average time range between the maximum and minimum values observed in the core total intensity light
curve:
\begin{equation} 
T_{\rm B,var}^{\rm obs}= 1.548 \times \, 10^{-32} \frac{\Delta S_{\rm max}d_L^2}{\nu^2\tau^2(1+z)^4} \ \ \ [\mbox{K}]
,\end{equation}
with $\Delta S_{\rm max}$ being the difference between the maximum and the minimum core flux density values and $d_L$ (m) the luminosity distance \citep[e.g.][]{Liodakis2017}. The $T_{\rm B,var}^{\rm obs}$ values at the three observing frequencies are reported in Table~\ref{tab_T_B} (fourth line).

\begin{table}
\centering
\caption{Jet PA values with standard errors ($\sigma_{PA}$), as reported in the bottom panel in Fig.~\ref{ridgelines}, and core region spectral indexes between 15 and 43\,GHz ($SI_{15--43}$) with the uncertainties $\sigma_{SI_{15--43}}$.}
\label{tab_PA_SI}
\begin{small}
\setlength{\tabcolsep}{4.5pt}
\renewcommand{\arraystretch}{1.0}
\begin{tabular}{lccccc}
\hline
\hline
Obs. date & MJD & PA & $\sigma_{PA}$ & $SI_{15-43}$ & $\sigma_{SI_{15-43}}$  \\
& &  (deg.) & (deg.) & & \\
\hline 
\vspace{-0.3cm} \\
2015/02/02      &       57055   &       56.2    &       1.1     &       0.21    &       0.13    \\
2015/02/25      &       57078   &       59.1    &       1.0     &       0.30    &       0.13    \\
2015/09/11      &       57276   &       54.4    &       0.5     &       0.28    &       0.13    \\
2015/10/17      &       57312   &       51.7    &       1.0     &       0.37    &       0.13    \\
2016/01/03      &       57390   &       45.1    &       0.7     &       0.16    &       0.13    \\
2016/02/20      &       57438   &       49.0    &       1.3     &       0.27    &       0.13    \\
2016/04/05      &       57483   &       53.7    &       1.0     &       0.35    &       0.13    \\
2016/05/17      &       57525   &       49.9    &       0.5     &       0.49    &       0.13    \\
2016/06/19      &       57558   &       39.4    &       1.2     &       0.47    &       0.13    \\
2016/09/19      &       57650   &       48.8    &       1.7     &       0.55    &       0.14    \\
2016/10/11      &       57672   &       49.8    &       1.6     &       0.42    &       0.13    \\
2017/01/02      &       57755   &       54.2    &       1.1     &       0.38    &       0.14    \\
2017/03/23      &       57835   &       49.5    &       1.2     &       0.28    &       0.13    \\
2017/06/12      &       57916   &       44.8    &       1.4     &       0.10    &       0.13    \\
\hline
\hline
\end{tabular} 
\end{small}
\end{table}

Since blazar jets are closely aligned to our line of sight, the effects of Doppler boosting must be taken into account when investigating the intrinsic source properties \citep{Urry1995}. 
$T_{\rm B,vlbi}^{\rm obs}$ and $T_{\rm B,var}^{\rm obs}$ are related to the intrinsic brightness temperature $T_{\rm B}^{\rm int}$ via the following equations:
 \begin{equation}
T_{\rm B,vlbi}^{\rm obs} = T_{\rm B}^{\rm int} \times \, \delta  \label{eq_T_vlbi}
 ,\end{equation}
 \begin{equation}
T_{\rm B,var}^{\rm obs} = T_{\rm B}^{\rm int} \times \, \delta^3 \label{eq_T_var}
 ,\end{equation}
where $\delta=[\gamma(1-\beta \cos\theta)]^{-1}$ is the Doppler factor (with $\theta$ being the viewing angle between the jet and our line of sight, $\beta$ the jet speed in units of the speed of light, and $\gamma=(1-\beta^2)^{-1/2}$ the bulk Lorentz factor).

\section{Discussion} \label{sec_discussion}
In recent years, several multi-frequency observing campaigns have been devoted to the study of the quasi-periodic variations (on timescales of $\sim2$ years) detected in the $\gamma$-ray light
curves of the PG\,1553+113 \citep{Ackermann2015}. 
While optical periodic variations on different timescales have been extensively investigated in blazars \citep[e.g.][]{Valtonen2006, Li2009, Britzen2018}, at $\gamma$-ray energies this has been possible only with the advent of the \fermi -LAT in 2008 thanks to the continuous and systematic sky monitoring in the MeV-GeV energy range. 
A similar approximately two-year quasi-periodicity (with significance above $>3\sigma$) has also been claimed in a few other blazars such as PKS 0537-441 \citep{Sandrinelli2016}, BL Lacertae \citep{Sandrinelli2017}, PKS 2155-304 \citep{Zhang2017a}, PKS 0301-243 \citep{Zhang2017b}, and J1043+2408 \citep{Bhatta2018}. 

\begin{figure}
\includegraphics[width=1.0\columnwidth, angle=90]{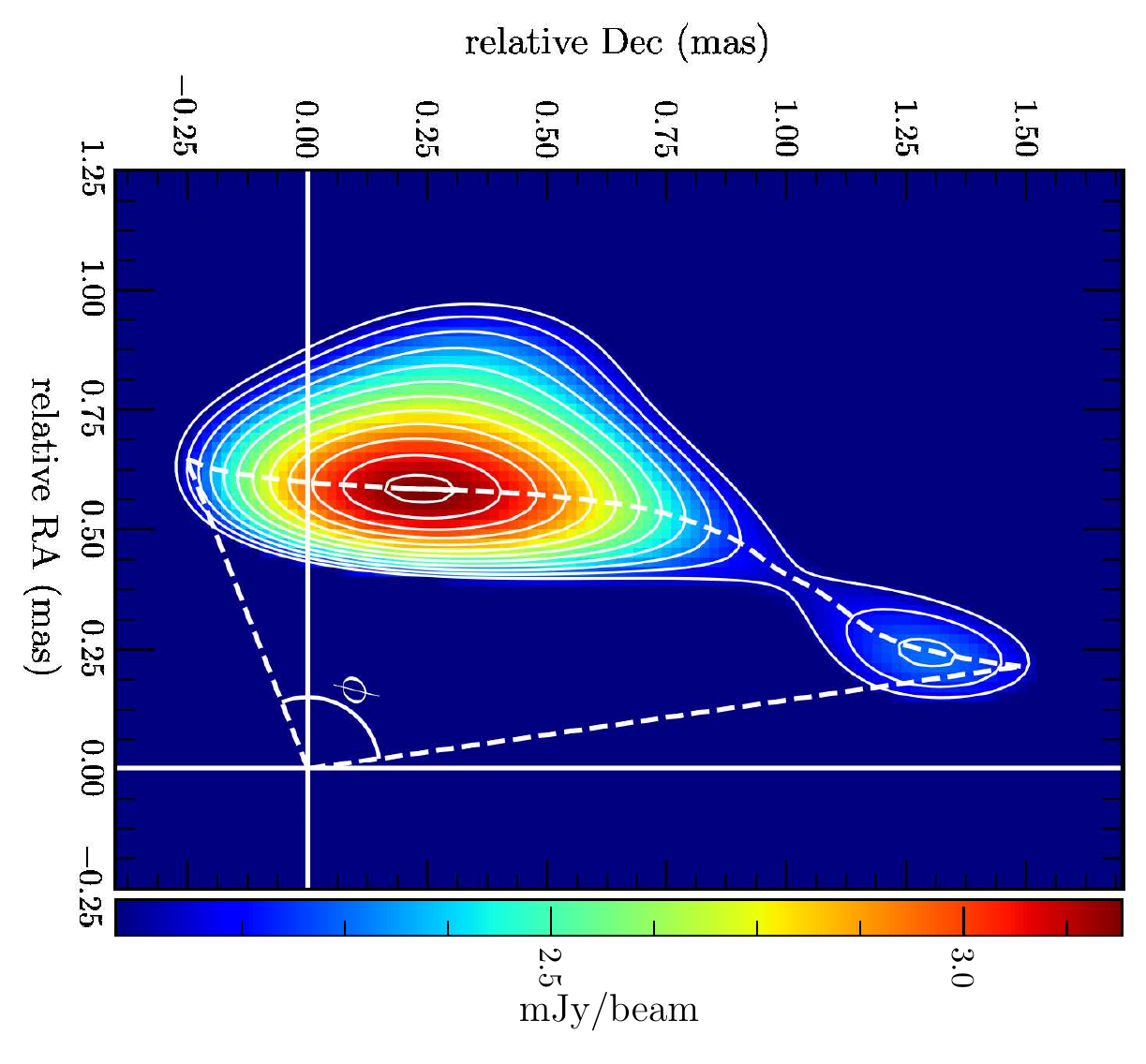}
 
\caption{The 15\,GHz residual stacked image after core subtraction (see Sect.~\ref{sec_ridgeline}). The colour scale represents the total intensity emission. The lowest contour is three times the image noise (that is $\sim0.34$ mJy/beam), with the following contours being a factor of two higher. The $\phi$ symbol represents the angle of the funnel in which the jet wobbles across time.}
\label{residuals}
\end{figure}

However, while some dedicated studies confirm the approximately two-year periodicity in the $\gamma$-ray light
curve of PG\,1553+113 \citep[e.g.][]{Ackermann2015, Prokhorov2017, AitBenkhali2019}, there are other studies which demonstrate that there is no solid evidence supporting such hints of year-long periodicities in blazars \citep[e.g.][]{Covino2019}.
We emphasise that caution is required when claiming quasi-periodicities on timescales of a few years in blazar $\gamma$-ray emission. This is because of the limited $\gamma$-ray light-curve duration (based on $\sim10$ years of LAT operational activity), and the red-noise contamination that can easily mimic such short-period variability patterns \citep[e.g.][]{Covino2019}.

One approach to further substantiate the existence of $\gamma$-ray periodicity is the use of comparisons to other wavelengths. In the case of PG\,1553+113, the optical light
curve shows a variability trend in good agreement with the $\gamma$ rays, confirming the approximately two-year periodic pattern. Moreover, hints of possible periodic behaviour are also found at X-rays, although this is not statistically significant due to the very sparse temporal sampling \citep{Ackermann2015, Tavani2018}. In this work we extend the investigation to the radio emission from the innermost jet region by means of a dedicated multi-frequency VLBA observing campaign.
It is apparent from the 15, 24, and 43\,GHz VLBA light
curves shown in Fig.~\ref{light_curves} that two periods of enhanced radio emission occur around October 2015 (MJD 57312) and June 2017 (MJD 57916), with a minimum appearing around June 2016 (MJD 57558). We note that the spectral index flattens during the periods of enhanced activity, with $\alpha_{15-43}\sim0.2\pm0.1$, while it is steeper during the low-activity state, with $\alpha_{15-43}\sim0.5\pm0.1$ (see Table~\ref{tab_PA_SI}). 
No clear repeating patterns are discernible in the light
curve during the period 2015--2017. Furthermore, the detection of multiple cycles on a longer observing period would be necessary to claim quasi-periodicity in the parsec-scale radio emission.

\begin{figure}
\includegraphics[width=1.0\columnwidth]{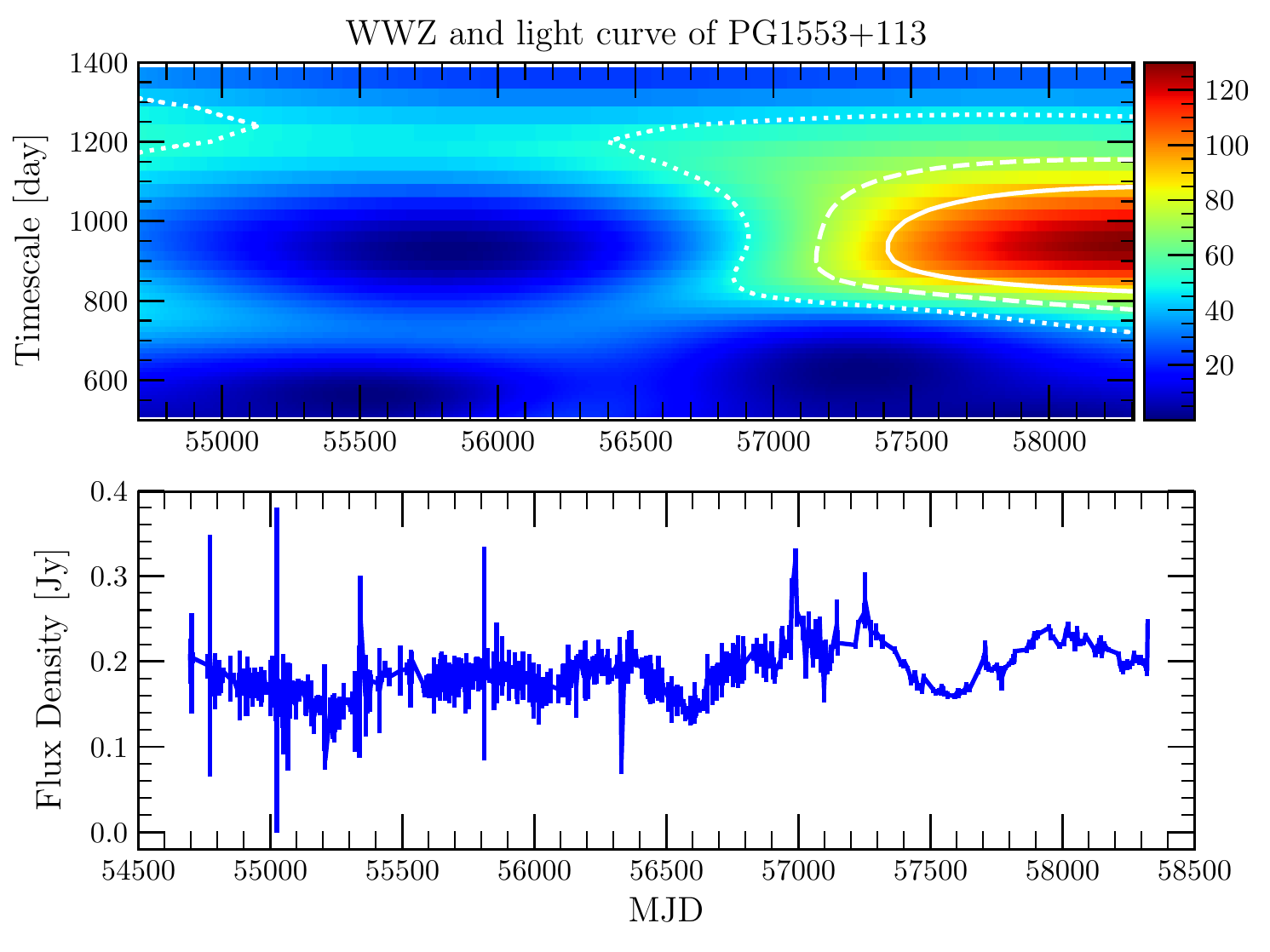}
    
\caption{Upper frame: Two-dimensional WWZ power (colour scale on the right side) distribution as a function of time (x-axis) and period (y-axis) of the 15\,GHz OVRO light
curve over the MJD range 54696-58153. The significance intervals are plotted as white contours, in which the dotted, dashed, and solid lines represent $1\sigma$, $2\sigma,$
and $3\sigma$, respectively. Lower frame: OVRO light
curve over the MJD range 54696-58153.} 
\label{wwz}
\end{figure}

We note that the 15\,GHz VLBA flux density trend over the entire observing period is in good agreement with the light
curve obtained from the OVRO monitoring (red triangles overlaid in the top panel of Fig.~\ref{light_curves}).
Indeed, we find $F_{var} \sim 0.13\pm0.03$ for both light
curves and the Pearson correlation coefficient between our 15\,GHz VLBA and the closest OVRO observations  (in time) is $r = 0.90$, implying a correlation above a $3\sigma$ level.
These findings allow us to argue that: (1) the overall variability in the radio emission is driven by changes occurring within the VLBI core region, and (2) that the OVRO light
curve is representative of the core region flux density modulation.
Based on these assumptions, it would appear that despite the presence of rapid variability in the OVRO light
curve since 2008 (lower frame in Fig.~\ref{wwz}) there is no clear and well-defined periodicity in the radio emission (in contrast to the well-defined periodicity in the $\gamma$-ray emission). 

This is confirmed by applying the weighted wavelet z-transform (WWZ) method \citep{Foster1996}, which is widely used for testing and quantifying quasi-periodic oscillations in blazar light
curves \citep[e.g.][]{Hovatta2008, Ackermann2015, Bhatta2017, Tavani2018, AitBenkhali2019}. 
We adopt a WWZ period search window in the frequency range $5.5 - 20.0 \times 10^{-4}$ day$^{-1}$, with a frequency step of $3 \times 10^{-5}$ day$^{-1}$, resulting in approximately $50$ test frequencies.
We choose the minimum frequency in such a way that at least two complete cycles could be detected throuth the full observing time range that covers $\sim10$ yr, while the maximum frequency corresponds to a  timescale of 500 days (i.e. about $80$ times the mean time separation). The wavelet window width is defined by using a decay constant c = 0.0125.
In Fig.~\ref{wwz} we show the OVRO light
curve WWZ power (colour scale in the top-right corner) as a function of time ($x-axis$) and period ($y-axis$) over the MJD range 54696-58153. 
From this analysis, no variability pattern is seen in the OVRO light
curve over the entire time range, in contrast to that seen in the $\gamma$-ray light
curve \citep{Ackermann2015, Tavani2018}.
A possible quasi-periodic modulation is found over a limited MJD range (57300-58153) with a period of $\sim930$ days  ($\sim2.5$ years). 

We assess the statistical significance of the WWZ variability timescale  by means of Monte Carlo simulations, following the approach proposed by \citet[][and references therein]{Emmanoulopoulos2013}, which takes into account the so-called `red noise leakage' and aliasing effects due to the sampling properties of the real data sets (i.e. finite length and finite time resolution).
We first generate 2000 light
curves with the same power spectrum density (PSD) and power density function (PDF) as the actual observations. We then calculate the WWZ periodograms for the simulated light
curves using the same parameters as for the OVRO 15\,GHz observations. The averaged WWZ is then compared with the one obtained from the observed light
curve. The significance intervals are over-plotted as white contours in Fig.~\ref{wwz}, in which the dotted, dashed, and solid lines represent $1\sigma$, $2\sigma,$ and $3\sigma$, respectively.

These results confirm the lack of a clear periodic variability or pattern that is stable over time at radio frequencies.

\begin{table}
\caption{Rotation measure and intrinsic polarisation angle values for the core region obtained from the linear fits of EVPAs vs. $\lambda^2$.}
\label{rm_table}     
\centering    
\tiny                 
\begin{tabular}{c c c c c c}        
\hline\hline
Epoch & MJD & RM\tablefootmark{a} & $\sigma_{RM}$\tablefootmark{b} & $\chi$\tablefootmark{c} & $\sigma_{\chi}$\tablefootmark{d} \\
mm/dd/yyyy & & (rad\,m$^{-2}$) & (rad\,m$^{-2}$) & (deg) & (deg) \\
\hline 
\vspace{-0.2cm} \\
02/02/2015      &       57055   &       $-12780$        &       435     &       165     &       6 \\
02/25/2015      &       57078   &       $-7940$ &       380     &       200     &       6 \\
09/19/2016      &       57650   &       $-10120$        &       430     &       210     &       7 \\
01/02/2017      &       57755   &       $-10630$        &       470     &       135     &       7 \\
\hline 
\hline
\end{tabular}
\tablefoot{
\begin{tiny}
\newline
\tablefoottext{a}{Rotation measure in rad\,m$^{-2}$;} \tablefoottext{b}{Rotation measure uncertainty;}\\
\tablefoottext{c}{Intrinsic polarisation angle in degrees;} \tablefoottext{d}{Intrinsic polarisation angle uncertainty.}\\
\end{tiny}
}
\end{table}

\subsection{A wobbling jet}
In general, quasi-periodic MWL emission modulation in blazars has one of two possible origins: (i) The first is a geometrical origin with periodic variations in the Doppler beaming factor produced by periodic changes of the angle between the jet axis and our line of sight. These changes in orientation may be related to the jet precessing or rotational motion, and/or helical structure within relativistic jets \citep[e.g.][]{Camenzind1992, Abraham2000, Stirling2003, Nakamura2004, Rieger2004, Raiteri2015}. (ii) The second is an intrinsic origin, with quasi-periodic plasma injection into the relativistic jet that is due to quasi-periodic instabilities in the accretion flow, producing the observed variability patterns in the emitted radiation \citep[e.g.][]{Honma1992, Tchekhovskoy2011, Pihajoki2013}. In the case of a geometrical effect, the quasi-periodic variability is only observed in the  frame of the observer, while in the case of pulsational instabilities in the accretion flow, such quasi-periodic emission variations are also present in the jet comoving frame.
In both scenarios the presence of a binary SMBH system is often invoked \citep[e.g.][]{Cavaliere2017, Sobacchi2017, Tavani2018, Caproni2017, Britzen2018}.
As suggested by \citet{Begelman1980}, binary SMBHs may produce wiggles, helical motion, and periodic variability in the jets of AGNs as a kinematical consequence of their orbital motion and of jet precession.

\begin{table}
\centering
\caption{Core region brightness temperature values in units of $10^{10}$\,K. For more details see Sect.~\ref{sec_Tb}.}
\label{tab_T_B}
\setlength{\tabcolsep}{8pt}
\renewcommand{\arraystretch}{1.5}
\begin{tabular}{lccc}
\hline
\hline
 & 15\,GHz & 24\,GHz &  43\,GHz \\
&($10^{10}$K)&($10^{10}$K)&($10^{10}$K) \\
\hline
$T_{\rm B,vlbi}^{\rm obs}(\rm max)$  &  3.4 & 2.0 & 0.7 \\
$T_{\rm B,vlbi}^{\rm obs}(\rm min)$   & 1.0 & 0.5 & 0.1 \\
$T_{\rm B,vlbi}^{\rm obs}(\rm avg)$     &       1.9 & 1.1 & 0.4 \\
$T_{\rm B,var}^{\rm obs}$       &       8.5 & 3.2 & 1.4 \\
\hline
\hline
\end{tabular} 
\end{table}

In order to test the jet precession scenario in PG\,1553+113, we constrain the jet position angle across epochs by means of the total intensity ridge line, as described in Sect.~\ref{sec_ridgeline}. 
The jet position angle, as shown in Fig.~\ref{ridgelines}, is found to be variable across the different epochs with a range of values between $\sim40^{\circ}$ and $\sim60^{\circ}$. 
We assessed the significance of the PA variability by means of a $\chi^2$ analysis, and the null hypothesis of a constant PA equal to the average value of $50^{\circ}$ could be rejected with a confidence level above $3\sigma$.
A direct signature of the  PA variation across epochs is the large angle ($\sim100^{\circ}$) of the funnel in which the jets wander. This was determined via stacking the  images of the 15\,GHz residuals for all the observing epochs (Fig.~\ref{residuals}). 
The wobbling jet motion indicates that geometrical effects can play a role in the observed emission variability through Doppler boosting modulation. 
However, we do not find any clear connection between the jet position angle variation and the total intensity emission or with the approximately two-year $\gamma$-ray quasi-periodic variability pattern. For this reason, under the assumption that $\gamma$ rays and radio emission are produced in the same region, we argue that the Doppler boosting modulation alone cannot account for the observed recurrent oscillations in the $\gamma$-ray light
curve. Additional physical mechanisms are required.
Within the framework of a binary SMBH system, one plausible scenario is that the secondary black hole crosses and perturbs the accretion disc of the primary black hole, inducing a temporary enhancement of the accretion rate, which in turn leads to increased jet emission \citep{Sillanpaa1988, Valtaoja2000, Caproni2017, Britzen2018}.
Another mechanism that could be responsible for the jet wobbling is the possible shuttle of the core closer to the jet apex, as reported in several studies \citep[e.g.][]{Kovalev2008, Niinuma2015}. This core-shuttle effect can result  from either changes in the physical conditions at the jet base or the ejection of a new component blended within the unresolved core region \citep[e.g.][]{Hodgson2017, Lisakov2017}.

One possible explanation for the presence of quasi-periodicity in $\gamma$ rays but not in radio could come from distinct emission regions between the two frequencies.
Within this context, an additional scenario for producing enhanced $\gamma$-ray emission is described by \citet{Bosch-Ramon2012} by means of the possible dynamical interaction of a relativistic jet with matter clumps or the atmosphere of an evolved star in the near vicinity of a SMBH. As the authors of this work point out, in general a considerably amount of dust, stars, and gas tend to cluster in the central region of galaxies and the relativistic jets in AGNs are expected to frequently interact with such ambient material.

\subsection{Polarisation properties}    
The polarised emission from the core region (lower image in Fig.~\ref{light_curves}) is on average more variable ($F_{var} \sim 0.40\pm0.05$) than the total intensity emission ($F_{var} \sim 0.20\pm0.03$), and no variability patterns can
be easily recognised during this 2015--2017 VLBA observing campaign. The most notable features in the polarised emission light
curves are represented by a $\sim5.0\pm0.6$\,mJy peak reached at all of the observing frequencies during the third observing epoch (MJD 57276), and a second one of $\sim3.5\pm0.4$\,mJy during the fourteenth observing epoch (MJD 57916). 
Both peaks in the polarised emission are detected during the two periods in which the source is undergoing enhanced activity in the total intensity emission. This behaviour may indicate that the processes producing the increased total intensity emission activity are also responsible for the enhanced polarised emission.

The Faraday corrected EVPAs at 15, 24, and 43\,GHz, as well as the intrinsic polarisation angle, obtained from the EVPAs versus $\lambda^2$ linear fits during MJD 57055, 57078, 57650, and 57755 have been found to vary between being roughly parallel to the  jet axis to being  roughly transverse, with no clear connection with the total intensity emission modulation. This indicates that the magnetic field configuration is also variable with time.
Such behaviour, together with the low degree of polarisation in the core region (on average $\sim2.5$ mJy), could be explained by the presence of multiple polarised subcomponents blended within the beam. The net observed polarisation properties integrated over the VLBI core region result from the sum of the relative contributions of each subcomponent, whose properties vary with time. 

The linear dependence found between the observed EVPAs with $\lambda^2$ indicate that the Faraday screen is mostly external to the emitting region \citep[e.g.][]{Burn1966}. The RM has been found to vary with time in a range between $-1.3$ and $-0.8\times \, 10^4$ rad \ m$^{-2}$. 
Given that the jet position angle wobbles across the different observing epochs, we can argue that the RM variations are produced by variations of the path length ($d_l$) through the external Faraday screen and/or variations in the electron density distribution $n_e$ \citep{Stirling2003}. 
The persistent negative sign in the observed RM across the different epochs, within the scenario in which  the Faraday screen is mostly represented by the hot accretion flow or wind outflow, could be explained by a misalignment of the jet axis with respect to the wind axis \citep{Park2019}.
If compared to Mrk\,421 \citep[e.g.][]{Lico2014} and Mrk\,501 \citep[e.g.][]{Hovatta2012}, the RM value found for PG\,1553+113 is the highest measured in a blazar of HSP type to date. 
However, sample sizes  are still not large enough  for a statistically significant characterisation of the average RM properties of the HSP blazar population. In this regard, the results of recent and ongoing MWL polarisation VLBA observations will be presented in a further publication.

\subsection{Brightness temperature}
By investigating the brightness temperature properties we can obtain important information regarding the physical conditions within the jet, namely the energy balance within the emitting region.
We find that the core brightness temperature decreases as the frequency increases, with $T_{\rm B,vlbi}^{\rm obs}(\rm 15\,GHz) > T_{\rm B,vlbi}^{\rm obs}(\rm 24\,GHz) > T_{\rm B,vlbi}^{\rm obs}(\rm 43\,GHz)$. 
According to theoretical models \citep[e.g.][]{Marscher1995}, there is a critical frequency $\nu_c $ below (above) which the brightness temperature increases (decreases) with frequency.
Within this scenario, and according to our findings, $\nu_c$ should be $<15$\,GHz. Furthermore, by investigating the brightness temperature in a large sample of radio jets in a range of frequencies between 2 and 86\,GHz, \citet{Lee2014} found on average $\nu_c \approx 9$\,GHz. 
This decreasing trend of $T_{\rm B,vlbi}^{\rm obs}$ with frequency could indicate that within the two emitting regions probed by the 15 and 43\,GHz VLBA observations there is an acceleration of the jet flow with increasing distance from the central engine \citep[see also ][]{Melia1989, Marscher1995, Lee2016, Nair2019}.

The variability brightness temperature $T_{\rm B,var}^{\rm obs}$ (fourth line in Table~\ref{tab_T_B}) for each observing frequency is higher than  $T_{\rm B,vlbi}^{\rm obs}$ , as expected because of the different dependence
on $\delta$. 
Given the limitations of the current data sets, we note that for estimating $T_{\rm B,var}^{\rm obs}$ the variability timescale is assumed to be of the order of the average time range between the maximum and minimum values observed in the core total intensity light
curve. Even though this approximation is not particularly accurate, it is suitable for the aims of the current analysis.
Using Eqs. \ref{eq_T_vlbi} and \ref{eq_T_var}, with $T_{\rm B,vlbi}^{\rm obs}$ at the peak flux density, we can estimate both  $T_{\rm B}^{\rm int}$ and $\delta$:
$$ T_{\rm B}^{\rm int} = \sqrt{\frac{T_{\rm B,vlbi}^{{\rm obs}^3}}{T_{\rm B,var}^{\rm obs}}} \;\;\;\;\;\; {\rm and} \;\;\;\;\;\; \delta= \sqrt{\frac{T_{\rm B,var}^{\rm obs}}{T_{\rm B,vlbi}^{\rm obs}}}.$$

We obtain an average value for $\delta$ of $\sim1.4$, which is a typical value for HSP objects as obtained from VLBI observations \citep[e.g.][]{Giroletti2004, Lico2012, Piner2018}. Such a low value for the Doppler factor further supports our conclusion that the Doppler boosting modulation alone cannot account for the observed variability.

The resulting $T_{\rm B}^{\rm int}$ is of the order of $1.5 \times \, 10^{10}$\,K, well below the $\sim 5 \times \, 10^{11} - 10^{12}$\,K inverse Compton limit \citep{Kellermann1969}, and in agreement with the typical $T_{B}^{obs}$ found in HSP blazars \citep[e.g.][]{Piner2010, Piner2014, Lico2016}.
The physical mechanism generally invoked for explaining the $T_B^{int}$ values in HSPs is energy equipartition between particles and the magnetic field. Invoking equipartition yields an upper limit of $\sim 5 \times \, 10^{10}$\,K \citep{Readhead1994}. 
The $T_B^{int}$ estimated from our 15, 24, and 43\,GHz observations is on average below the equipartition limit, indicating that we are probing an emitting region where the magnetic field energy density ($u_B$) is higher than the non-thermal particle energy density ($u_p$). Using equation (5d) from \citet{Readhead1994} we find log($u_p/u_B) \sim -4.6$. A similar physical scenario is found in the innermost regions of the radio galaxy M\,87, as reported in \citet{Kim2018}.

\section{Summary and conclusions}  \label{sec_conclusion}
The HSP blazar PG\,1553+113 has been observed intensely since an approximately two-year quasi-periodic variability pattern was recognised in its $\gamma$-ray light
curve \citep{Ackermann2015}. 
In this work we explored the parsec-scale radio properties of the source by means of a 15, 24, and 43\,GHz VLBA observing campaign in total and polarised intensity during the period 2015--2017.

Two periods of enhanced activity around October 2015 and June 2017 emerge from the total intensity light
curve with a minimum around June 2016. However, in contrast to the $\gamma$-ray emission, no hints of quasi-periodic variability are found in the VLBI emission or in the 15\,GHz OVRO light
curve over a period covering about 10 years.

We detect polarised emission in the core region (with a polarisation percentage varying in the range $\sim1-4\%$) that is variable across epochs with no clear recurrent patterns in the light
curve. The polarisation angle has been found to be variable across epochs as well, but without any clear connection to the total intensity or polarised emission. We also find a variable RM in the core region, ranging between $-1.3$ and $-0.8\times \, 10^4$ rad \ m$^{-2}$.

The core brightness temperature is found to decrease with increasing frequency, in agreement with \citet{Lee2016}, likely suggesting that an acceleration of the jet flow is occurring within the emitting regions probed by the 15 and 43\,GHz VLBA observations \citep{Melia1989, Marscher1995}. Using both $T_{\rm B,vlbi}^{\rm obs}$ and $T_{\rm B,var}^{\rm obs}$ we estimate a Doppler factor of $\sim1.4$ and  $T_{\rm B}^{\rm int} \sim 1.5 \times \, 10^{10}$\,K, indicating that within the emitting region the total energy is dominated by the magnetic field. 

By means of a total intensity ridge-line analysis we constrain the jet position angle across the different observing epochs. We find that the jet direction varies in range between $\sim40^{\circ}$ and $\sim60^{\circ}$, indicating that geometric effects could play a role in the observed emission variability through Doppler boosting modulation. 
However, there is no direct and clear connection between the jet wobbling motion and either the radio flux density or the $\gamma$-ray variability pattern, and additional physical mechanisms are invoked. One possible mechanism responsible for the jet wobbling is the core-shuttle effect described by \citet{Hodgson2017} and \citet{Lisakov2017}. Another possibility is the presence of a binary SMBH system at subparsec separation, inducing a precessing motion in the jet as well as perturbations in the accretion disc with a consequent modulation in the accretion rate \citep[e.g.][]{Ackermann2015, Caproni2017, Tavani2018}. 
While evidence of binary SMBHs systems at kiloparsec scales have been found in a few tens of objects through direct detection of the two centres, parsec or subparsec systems, which are expected to form quickly according to evolution models, are more elusive. 
The detection of parsec-
or subparsec-scale binary SMBH systems would also be an important assessment of the prediction of coalescence of SMBHs and of consequent gravitational wave production \citep{Begelman1980, Bhatta2018, AitBenkhali2019}. 
This goal could be achieved in the next few years thanks to the efforts of the Event Horizon Telescope project \citep{EHT2019}, which could potentially allow us to spatially resolve sub-parsec binary SMBH systems.

With the present VLBA monitoring we provide new and valuable information for MWL studies of this peculiar object, furthering our understanding of the physical mechanisms that produce the observed periodicity in the $\gamma$-ray emission.
This work is intended to be part of a wider and extensive MWL observing program with regular monitoring of the source since 2015 at different frequencies, the results of which will be presented in a dedicated paper (MAGIC Coll. et al., in prep).

\begin{acknowledgements} \label{aknowlwdgements}
We thank the anonymous referee for carefully revising the paper and for the valuable and constructive comments that improved the paper.
RL gratefully acknowledges the financial support and the kind hospitality from the Instituto de Astrof\'{\i}sica de Andalucia (IAA-CSIC) in Granada. We that Jae-Young Kim for reading the paper and for the fruitful discussion. JL acknowledges the following grants: National Key R\&D Program of China (No.\,2018YFA0404602), Light in China’s Western Region program (No.\,2015-XBQN-B-01), National Natural Science Foundation of China (No.\,11503071). This work is based on observations obtained through the BL214 and BL215 VLBA projects, which makes use of the Swinburne University of Technology software correlator, developed as part of the Australian Major National Research Facilities Programme and operated under license \citep{Deller2011}. The National Radio Astronomy Observatory is a facility of the National Science Foundation operated under cooperative agreement by Associated Universities, Inc. 
This research has made use of data from the OVRO 40-m monitoring program \citep{Richards2011} which is supported in part by NASA grants NNX08AW31G, NNX11A043G, and NNX14AQ89G and NSF grants AST-0808050 and AST-1109911.
\end{acknowledgements}

\end{document}